\documentclass[a4paper,11pt]{article}
\usepackage{graphicx}
\usepackage{amssymb}

\def\arxiv#1{\texttt{#1}}
\def\mo{\stackrel{\circ}{M}}
\begin{document}

\begin{flushright}
\today\\
LPT-ORSAY/07-12
\end{flushright}

\vspace{0.3cm}

\begin{center}
{\bf {\Large Low-energy $\pi\pi$ and $\pi K$ scatterings revisited\\
in three-flavour resummed chiral perturbation theory}}\footnote{
Work supported in part by the EU Contract No. MRTN-CT-2006-035482, \lq\lq FLAVIAnet''.}

\vspace{0.5cm}

{\Large S. Descotes-Genon}

\vspace{0.3cm}

\emph{Laboratoire de Physique Th\'eorique,\\
CNRS/Univ. Paris-Sud 11 (UMR 8627), 91405 Orsay Cedex, France}

\end{center}

\vspace{0.5cm}

\begin{center}

{\bf Abstract}

\end{center}

Chiral symmetry breaking may exhibit significantly different patterns in two chiral limits: $N_f=2$ massless flavours ($m_u=m_d=0$, $m_s$ physical) and $N_f=3$ massless flavours ($m_u=m_d=m_s=0$).
Such a difference may arise due to vacuum fluctuations of $s\bar{s}$ pairs related to the
violation of the Zweig rule in the scalar sector, and could yield a numerical competition 
between contributions counted as leading and next-to-leading order in the chiral expansions
of observables. We recall and extend Resummed Chiral Perturbation Theory (Re$\chi$PT), 
a framework that we introduced previously to deal with
such instabilities: it requires a more careful definition of the relevant 
observables and their one-loop chiral expansions. We
analyse the amplitudes for low-energy $\pi\pi$ and $\pi K$ scatterings within
Re$\chi$PT, which we match in subthreshold regions with dispersive representations 
obtained from the solutions of Roy and Roy-Steiner equations.
Using a frequentist approach, we constrain
the quark mass ratio as well as the quark 
condensate and  the pseudoscalar decay
constant in the $N_f=3$ chiral limit. The results mildly favour significant
contributions of vacuum fluctuations suppressing the $N_f=3$ quark condensate
 compared to its $N_f=2$ counterpart. 

\newpage

\section{Introduction}

A striking feature of the Standard Model consists in the 
mass hierarchy obeyed by the light quarks:
\begin{equation}
m_u\sim m_d \ll m_s \sim \Lambda_{QCD} \ll \Lambda_H\,,
\end{equation}
where $\Lambda_{QCD}$ is the characteristic scale describing the
running of the QCD effective coupling and $\Lambda_H\sim 1$ GeV the mass
scale of the bound states not protected by chiral symmetry. Therefore,
the strange quark may play a special role in the low-energy dynamics of
QCD: 

\emph{i)} it is light enough to allow for a combined expansion of
observables in powers of $m_u,m_d,m_s$ around the $N_f=3$
chiral limit (meaning 3 massless flavours):
\begin{equation}
N_f=3:\qquad m_u=m_d=m_s=0\,,
\end{equation}

\emph{ii)} it is sufficiently heavy
to induce significant changes in order parameters
from the $N_f=3$ chiral limit to
the $N_f=2$ chiral limit (meaning 2 massless flavours):
\begin{equation}
N_f=2:\qquad m_u=m_d=0 \qquad m_s {\rm\ physical}\,,
\end{equation}

\emph{iii)} it is too light to suppress efficiently
loop effects of massive $\bar{s}s$ pairs (contrary to $c,b,t$ quarks).

These three arguments suggest that $\bar{s}s$ sea pairs may play
a significant role in chiral dynamics, leading to different patterns of chiral symmetry
breaking in $N_f=2$ and $N_f=3$ chiral limits. Then, chiral order parameters
such as the quark condensate and the pseudoscalar decay constant:
\begin{equation} \label{eq:chirorddef}
\Sigma(N_f)=-\lim_{N_f} \langle\bar{u}u\rangle\,, \qquad \qquad
F^2(N_f)=\lim_{N_f} F^2_\pi\,,
\end{equation}
would have significantly different values in the two chiral limits 
($\lim_{N_f}$ denoting the chiral limit with $N_f$ massless flavours). 

The role of $\bar{s}s$-pairs in the structure of QCD vacuum
is a typical loop effect : it should be suppressed in 
the large-$N_c$ limit, and it can be significant
only if the Zweig rule is badly violated in the vacuum (scalar) channel
$J^{PC}=0^{++}$. On general theoretical
grounds~\cite{param}, one expects $\bar{s}s$ sea-quark pairs to have a 
paramagnetic effect on chiral order parameters, so that they should
decrease when the strange quark mass is sent to zero : for instance,
$\Sigma(2;m_s)\geq \Sigma(2;m_s=0)$, and similarly for $F^2$, which
translates into the paramagnetic inequalities:
\begin{equation} \label{eq:paramag}
\Sigma(2) \geq \Sigma(3)\,, \qquad F^2(2) \geq  F^2(3)\,.
\end{equation}
However, the size of this paramagnetic suppression is not predicted. Thus, it is highly desirable to extract
the size of the chiral order parameters in $N_f=2$ and $N_f=3$ limits
from experiment.

Recent data on $\pi\pi$ scattering~\cite{E865} together with older data
and numerical solutions of the Roy 
equations~\cite{ACGL} allowed us to determine
the $N_f=2$ order parameters expressed in suitable physical 
units~\cite{pipi}:
\begin{eqnarray} \label{eq:x2}
X(2)&=&\frac{(m_u+m_d)\Sigma(2)}{F_\pi^2M_\pi^2}=0.81\pm 0.07\,,\\
\label{eq:z2}
Z(2)&=&\frac{F^2(2)}{F_\pi^2}=0.89\pm 0.03\,.
\end{eqnarray}
A different analysis of the data in ref.~\cite{E865}, with
the additional input of dispersive estimates for the (non-strange) scalar radius
of the pion, led to a larger value of $X(2)$~\cite{CGL}. 
$X(2)$ and $Z(2)$ seem fairly close to 1, so that corrections related to
$m_u,m_d\neq 0$ (while $m_s$ remains at its physical value) have a limited
impact on the low-energy behaviour of QCD. In turn, 
two-flavour Chiral Perturbation Theory ($\chi$PT)~\cite{chpt2},
which consists in an expansion in powers of $m_u$ and $m_d$
around the $N_f=2$ chiral limit, would not suffer from severe
problems of convergence~\footnote{Let us stress that
new data of high accuracy are expected from the NA48/2 
collaboration soon, which could affect these results 
significantly~\cite{NA48}. Recent lattice simulations with two-flavour 
dynamical quarks~\cite{lattice} may help to understand some aspects of these questions,
even though the results are preliminary and rather delicate to
interpret.}. 

Two-flavour $\chi$PT~\cite{chpt2}  
deals only with dynamical pions in a very limited range of energy. 
In order to include $K$- and $\eta$-mesons dynamically
and extend the energy range of interest, one must use
three-flavour $\chi$PT~\cite{chpt3}
where the expansion in the three light-quark masses starts 
around the $N_f=3$ vacuum $m_u=m_d=m_s=0$.
From the above discussion, large vacuum fluctuations of $\bar{s}s$
pairs would have a dramatic effect on $N_f=3$ chiral expansions.
The leading-order (LO) term, which depends on the $O(p^2)$ low-energy
constants $F^2(3)$ and $\Sigma(3)$, would be damped. On the other
hand, next-to-leading-order (NLO) corrections could be enhanced, in
particular those related to Zweig-rule violation in the scalar sector.
For instance, the Gell-Mann--Oakes--Renner relation would not be
saturated by its LO term and would receive sizable numerical
contributions from terms counted as NLO in the chiral counting.

Unfortunately, the experimental data on $K$- and $\eta$-decays are not
accurate enough to assess the role of $s\bar{s}$ pairs in the
$N_f$-dependence of chiral symmetry breaking in a very precise way. 
However our understanding
of $\pi K$ scattering at low energies has been improved recently 
through the re-analysis of dispersive Roy-Steiner equations~\cite{roypika}. A 
rapid analysis of its results in the framework of three-flavour $\chi$PT hinted
at significant vacuum fluctuations encoded in some $O(p^4)$ chiral couplings, 
which calls for a more detailed analysis of the $\pi K$ system. Interesting
information can also be obtained from our current knowledge of $\pi\pi$
scattering, which we will include in our study. 
 
To perform such an analysis, we develop and modify the framework presented 
in refs.~\cite{param,ordfluc,resum}. Specifically, our work differs from
ref.~\cite{resum} on three points: we consider not only $\pi\pi$- but also $\pi
K$-scattering, our observables are the values of the amplitudes in unphysical
 regions rather than subtraction constants of dispersion relations, 
the matching between theoretical and experimental representations is performed
in a frequentist approach, not in a Bayesian framework.

In sec.~2, we motivate and explain Resummed Chiral Perturbation Theory (Re$\chi$PT), a
framework designed to derive three-flavour chiral
series at one loop, in which vacuum fluctuations of $s\bar{s}$ pairs are
resummed. In sec.~3, we apply 
Re$\chi$PT to $\pi\pi$- and $\pi K$-scattering amplitudes.
In sec.~4, we explain how we determine the same amplitudes dispersively in
subthreshold regions, building upon the solutions of Roy and Roy-Steiner equations~\cite{ACGL,roypika}.
In sec.~5, we discuss the matching of the chiral and dispersive results within a frequentist
approach~\cite{rfit}, and in sec.~6, 
we present our results for the order parameters of $N_f=3$ chiral symmetry breaking.
In sec.~7, we summarise and discuss our results. Appendices are devoted to the
expression of scattering amplitudes in Re$\chi$PT,
their evaluation from Roy and Roy-Steiner equations and the
treatment of correlated data.

\section{Resummed Chiral Perturbation Theory}

We start by describing in more detail the framework introduced in
refs.~\cite{param,ordfluc,resum} to expand observables around the $N_f=3$
chiral limit in the case of significant vacuum fluctuations. We take this
opportunity to extend this framework to deal with energy-dependent quantities.

\subsection{Convergence of observables}

In the introduction, we have emphasised the possibility for 
three-flavour chiral series to exhibit a rather unusual behaviour,
with a numerical competition between leading and next-to-leading
order. In ref.~\cite{resum}, we 
called instability of the expansion such a numerical 
competition between terms of different chiral counting.
A na\"{\i}ve argument based on resonance saturation suggests that
higher orders in the chiral expansion should be suppressed by powers 
of $(F_\pi/\Lambda_H)^2$. However, such an argument does not apply
to a leading-order contribution proportional to $\Sigma(3)$~\cite{gchpt}: there is no 
resonance that could saturate the quark condensate. Therefore we expect
to encounter three-flavour chiral expansions with a good overall convergence:
\begin{equation} \label{eq:conv}
A=A_{LO} + A_{NLO} + A\delta A\,, \qquad\qquad  \delta A\ll 1\,,
\end{equation}
but the numerical balance between the leading order $A_{LO}$ and 
the next-to-leading order $A_{NLO}$ depends on 
the importance of vacuum fluctuations of $s\bar{s}$ pairs. 

At the level of $O(p^4)$ $N_f=3$ chiral perturbation theory,
the size of the vacuum fluctuations is encoded in the low-energy constants
(LECs) $L_4$ and $L_6$ whose values remain largely unknown. For a long
time~\cite{chpt3}, they have been set to zero at an arbitrary hadronic scale 
(typically the $\eta$-mass)
assuming that the Zweig rule held in the scalar sector. 
More recent but indirect analyses based on dispersive 
methods~\cite{roypika,dispff,dispzr} 
suggest values of $L_4$ and $L_6$ which look quite modest but are sufficient
to drive the $N_f=3$ order parameters $\Sigma(3)$ and $F^2(3)$ 
down to half of their $N_f=2$ counterparts $\Sigma(2)$ and
$F^2(2)$, leading to $A_{LO}\simeq A_{NLO}$ 
as recalled in sec.~\ref{sec:massdec}. In addition, two-loop 
analyses~\cite{twoloopscal,twolooppipi,twolooppika,twolooprev} 
led to values of $L_4$ and $L_6$ off large-$N_c$ expectations.

Unstable $N_f=3$ chiral expansions ($A_{LO}\sim A_{NLO}$)
demand a more careful treatment than in two-flavour $\chi$PT where
such instabilities are seemingly absent. For instance, it would be wrong 
to believe that the chiral expansion of $1/A$
converges nicely~\footnote{This would be equivalent to claiming that
$1/(1+x)\simeq 1-x$ is a reasonable approximation for $x=O(1)$.}.
This might induce the observed problems of convergence in recent two-loop 
computations~\cite{twoloopscal,twolooppipi,twolooppika}~: 
the latter treat the fluctuations encoded
in $L_4$ and $L_6$ as small and are not designed to 
cope with a large violation of the Zweig rule in the scalar sector,
leading to instabilities of the chiral series.

Observables with a good convergence in the sense of eq.~(\ref{eq:conv})
form a linear space, which we identify with connected QCD correlators of
axial/vector currents and their derivatives, away from kinematic
singularities. This choice promotes some ``good'' observables that can 
be extracted from such correlators, such as 
$F_P^2$ and $F_P^2M_P^2$ ($P=\pi,K,\eta$) : LO and NLO may compete, 
but there should be only a tiny contribution from NNLO and higher orders. 
On the contrary, the chiral expansion of $M_P^2$ 
(ratio of the former quantities) may exhibit a bad convergence. Similarly,
the good observable associated with a form factor $F_{P\to Q}$ describing 
a transition from a pseudoscalar meson $P$ to a meson $Q$ 
will be $F_P F_Q F_{P \to Q}$, where
the decay constants $F_P$ and $F_Q$ stem from wave-function
renormalisation factors in the LSZ reduction formula.

\subsection{One-loop bare expansion of QCD Green functions}\label{sec:rechipt}

In a previous work~\cite{resum}, we proposed 
a framework to deal with chiral expansions in the case of large
fluctuations, by resumming the terms 
containing the Zweig-rule violating LECs $L_4$ and $L_6$. 
This framework, which we will call Resummed Chiral Perturbation Theory
(Re$\chi$PT), includes consistently the
alternatives of large and small vacuum fluctuations. In this section,
we explain how to expand a good observable at one loop in Re$\chi$PT. We 
addressed only energy-independent quantities in ref.~\cite{resum}, where
we explained in detail the similarities and differences of our approach
with respect to Generalized Chiral Perturbation Theory~\cite{gchpt,KMSF}.

We start from the one-loop generating functional for three-flavour $\chi$PT~\cite{chpt3}:
\begin{equation} \label{eq:oneloop-gen}
Z=Z_t+Z_u+Z_A + \ldots
\end{equation}
where the ellipsis stands for NNLO contributions. The three terms of the
one-loop generating functional are:
\begin{itemize}
\item $Z_t$ is the sum of $O(p^2)$ and $O(p^4)$ tree graphs, and of 
tadpole contributions:
\begin{eqnarray} \label{zt}
Z_t&=&\sum_P\int dx 
   \frac{F_0^2}{6}\left\{1-\frac{3}{16\pi^2}\frac{\mo_P^2}{F_0^2}\log\frac{\mo_P^2}{\mu^2}\right\}
         \sigma^\Delta_{PP}\\ \nonumber
 &&\quad  + \sum_P\int dx 
   \frac{3F_0^2}{6}\left\{1-\frac{3}{6\pi^2}\frac{\mo_P^2}{F_0^2}\log\frac{\mo_P^2}{\mu^2}\right\}
         \sigma^\chi_{PP} + \int dx\ {\mathcal L}_4^r\,,
\end{eqnarray}
where $F_0 \equiv F(3)$, 
\begin{equation}
F_0 = F(3)\,, \qquad B_0=\frac{\Sigma(3)}{F(3)^2}\qquad\,,
r=\frac{m_s}{m}
\end{equation}
$\sigma^\Delta$ and $\sigma^\chi$ collect source terms for vector / axial currents
and scalar / pseudoscalar densities, and 
${\mathcal L}_4^r$ is the $O(p^4)$ chiral Lagrangian
with renormalised couplings $L_i^r$ and $H_i^r$.
$\mo_P^2$ denotes the $O(p^2)$ contribution to the (squared) mass
of the Goldstone boson $P$:
\begin{equation}\label{eq:truncmass}
\mo_\pi^2 = Y(3)M_\pi^2\,, \qquad \mo_K^2 = \frac{r+1}{2}Y(3)M_\pi^2\,,
\qquad \mo_\eta^2 = \frac{2r+1}{3}Y(3)M_\pi^2\,.
\end{equation}

\item $Z_u$ collects unitarity corrections 
corresponding to one-loop graphs with two $O(p^2)$ vertices:
\begin{eqnarray} \label{zu}
Z_u&=& \sum_{P,Q} 
  \int dx\ dy \Big[\{\{\partial_{\mu\nu}-g_{\mu\nu}\Box \}M_{PQ}^r(x-y)-g_{\mu\nu}L_{PQ}(x-y)\} 
         \hat\Gamma^\mu_{PQ}(x)\hat\Gamma^\nu_{QP}(y)\\ \nonumber
 &&\quad -\partial_\mu K_{PQ}(x-y) \hat\Gamma^\mu_{PQ}(x)\bar\sigma_{QP}(y)
         +\frac{1}{4} J^r(x-y)  \bar\sigma_{PQ}(x)\bar\sigma_{QP}(y)\Big]
\end{eqnarray}
where $J,K,L,M$ are (renormalised) functions defined from the one-loop scalar
integral with mesons $P$ and $Q$ propagating in the loop, 
and $\hat\Gamma^\mu$ and $\bar\sigma=\sigma^\Delta+\sigma^\chi$ collect source terms.
\item $Z_A$ is the Wess-Zumino functional collecting anomalous contributions.
\end{itemize}

The one-loop functional eq.~(\ref{eq:oneloop-gen}) has been derived using
the propagators and couplings of the $O(p^2)$ chiral Lagrangian, and therefore it is expressed only
in terms of chiral couplings: $F_0$ and $B_0$, $L_i$\ldots~\cite{chpt3}
In particular, the Goldstone degrees of freedom 
have masses truncated at $O(p^2)$, denoted $\mo_P^2$. Large
fluctuations should induce significant differences between this quantity
and the physical mass $M_P^2$. Therefore, we want to 
replace $\mo_P^2$ by $M_P^2$ only when justified by physics arguments, 
since this replacement may have an important impact when comparing chiral
expansions with experimental data.
\begin{itemize}
\item The anomalous contribution $Z_A$ corresponds to local couplings for vector and axial currents, and is not affected by our discussion.
\item For the unitarity corrections $Z_u$, were we to consider higher and higher orders of the chiral 
expansion, we should obtain that the masses occurring in the functions $J^r$,
$K$, $L$ and $M^r$  are physical masses, in order to get the low-mass two-particle cuts 
at the physical positions. Therefore, we write those functions 
with the physical masses of the Goldstone bosons. On the contrary, we keep the
multiplying factors $\Gamma^\mu$ and $\bar\sigma$ expressed in terms of parameters
of the effective Lagrangian ($m_q$, $B_0$\ldots).
\item The tadpole contributions present in $Z_t$ are derived using the $O(p^2)$ contribution
to the Goldstone boson masses $\mo_P^2$. In ref.~\cite{resum}, we have proposed the replacement:
\begin{equation} \label{eq:tadpole}
\frac{\mo_P^2}{32\pi^2}\log\frac{\mo_P^2}{\mu^2} \quad \to \quad 
\frac{\mo_P^2}{32\pi^2}\log\frac{M_P^2}{\mu^2}\,.
\end{equation}
We could have kept $\mo_P^2$ everywhere in $Z_t$, and in particular inside the logarithm. However,
the resulting expressions are easier to deal with, and the change has only a tiny numerical impact: 
either $M_P$ is close to its $O(p^2)$ term and the change is trivially
justified, or $\mo_P^2$ is much smaller than $M_P^2$ and the whole
tadpole contribution is very small. 
\item Physical $S$-matrix elements are obtained from
the Green functions derived with the generating functional by
applying the LSZ reduction formula. The external legs corresponding
to incoming and outgoing particles must be put on the mass shell.
In the process, the products of external momenta are translated into 
the well-known Mandelstam variables. These kinematical relations are 
valid for physical masses, and we will use the
latter (and not the $O(p^2)$ truncated masses $\mo_P^2$)
whenever we reexpress products of external momenta.
This prescription is consistent with the use of physical masses in the
one-loop scalar integral present in the unitarity term $Z_u$.
\end{itemize}

Following the renormalisation procedure in ref.~\cite{chpt3},
one can check easily that eq.~(\ref{eq:tadpole}) does not change the
renormalisation-scale dependence of LECs at one loop. Actually, 
the whole one-loop generating functional $Z$ becomes exactly
renormalisation-scale independent : when we follow the prescription given above,
all the scale-dependent logarithms present in $Z_t$ (explictly shown in eq.~(\ref{zt})) and 
$Z_u$ (hidden in the one-loop functions $M^r$ and $J^r$ in eq.~(\ref{zu})) 
are multiplied by terms of the same form $m_q B_0$ and thus cancel exactly. In the more
usual treatment of the tadpoles~\cite{chpt3}, $m_q B_0$ terms are replaced
by physical Goldstone masses in the one-loop generating functional (see sec.~8 in
ref.~\cite{chpt3}). In this case, the cancellation of the logarithms takes place only up to $O(p^4)$
and some higher-order logarithmic pieces of $Z_t$ have no counterpart in $Z_u$.

We call ``bare expansion'' the chiral expansion treated according to our prescription,
because of we prefer keeping original couplings of the chiral Lagrangian
to trading them for physical masses and decay constants.
We sum up our method to obtain bare expansions of Green functions in Resummed $\chi$PT:
\begin{enumerate}
\item Consider a subset of observables suitable for a chiral expansion, such as
the linear space of connected QCD correlators of axial/vector currents and their derivatives away 
from kinematic singularities.
\item Extract the bare expansion of the observables using the 
one-loop generating functional eq.~(\ref{eq:oneloop-gen}): in $Z_t$, replace
the tadpole contributions by 
eq.~(\ref{eq:tadpole}), and in $Z_u$, use the  
physical masses for the functions $J,K,L,M$ defined from the one-loop scalar integral.
\item Use physical masses to reexpress scalar products of external momenta
in terms of the Mandelstam variables.
\item Keep track of the higher-order contributions by introducing
remainders, i.e. NNLO quantities which have an unknown value
but are assumed small enough for the chiral series to converge.
\item Exploit algebraically the resulting relations, and never trade the 
couplings of the chiral Lagrangian for observables
while neglecting higher-order terms.
\end{enumerate}
The main differences from the usual treatment of three-flavour chiral series consists in
the choice of a particular subset of observables, the distinction between physical meson masses and their
$O(p^2)$ truncated forms, and the algebraic use of chiral expansions 
while keeping track of higher-order terms explicitly.

\subsection{Masses and decay constants of Goldstone bosons} \label{sec:massdec}

The first example consists in pseudoscalar decay constants and
masses. The usual $\chi$PT expressions (Sec.~10 in ref.~\cite{chpt3})
become the following bare expansions in Re$\chi$PT 
(similar expressions for $\eta$ can be found in refs.~\cite{resum,zr}):
\begin{eqnarray} \label{fpi}
F_\pi^2 &=& F_\pi^2 Z(3)
  +8(r+2)Y(3)M_\pi^2 \Delta L_4
  +8 Y(3)M_\pi^2 \Delta L_5+F_\pi^2e_\pi\,,\\
\label{fka}
F_K^2 &=& F_\pi^2 Z(3)
  +8(r+2)Y(3)M_\pi^2 \Delta L_4
  +4(r+1)Y(3)M_\pi^2 \Delta L_5+F_K^2e_K\,,\\
\label{fpimpi}
F_\pi^2M_\pi^2 &=& F_\pi^2M_\pi^2 X(3)
  +16(r+2)Y^2(3)M_\pi^4 \Delta L_6
  +16 Y^2(3)M_\pi^4 \Delta L_8+F_\pi^2M_\pi^2d_\pi\,,\\
\label{fkamka}
F_K^2M_K^2 &=& \frac{r+1}{2} F_\pi^2M_\pi^2 X(3)\\ \nonumber
&&\qquad  +8(r+2)(r+1)Y^2(3)M_\pi^4 \Delta L_6
  +4(r+1)^2Y^2(3)M_\pi^4 \Delta L_8+F_K^2M_K^2d_K\,,
\end{eqnarray}
We take as free parameters the $N_f=3$ quark condensate and pseudoscalar decay constant
  expressed in physical units~\footnote{In this paper, we work in the 
isospin symmetry limit, where $m_u=m_d=m$ and the electromagnetic interaction
is ignored. We take the following values for
the masses and decay constants: $F_\pi=92.4$ MeV, $F_K/F_\pi=1.22$,
$M_\pi=139.6$ MeV, $M_K=495.7$ MeV, $M_\eta=547$ MeV.}, as well as their ratio
  and the quark mass ratio:
\begin{equation}
X(3)=\frac{2m\Sigma(3)}{F_\pi^2 M_\pi^2}\,, \quad 
Z(3)=\frac{F^2(3)}{F_\pi^2}\,, \quad
Y(3)=\frac{X(3)}{Z(3)}=\frac{2mB_0}{M_\pi^2}\,,
r=\frac{m_s}{m}
\end{equation}
We have introduced the NNLO remainders 
$d_\pi$, $e_\pi$, $d_K$ and $e_K$, and
the combinations of LECs and chiral logarithms:
\begin{eqnarray}
\Delta L_4
       &=& L_4^r(\mu) -\frac{1}{256\pi^2}
          \log\frac{M_K^2}{\mu^2} \nonumber\\
 &&\quad
          +\frac{1}{128 \pi^2} \frac{r}{(r-1)(r+2)} \left\{ \log \frac{M_K^2}{M_\pi^2}
           +\left(1 + \frac{1}{2 r} \right) \log
          \frac{M_\eta^2}{M_\pi^2}  \right\} \label{deltal4} \,,\\
 \Delta L_5 &=& L_5^r(\mu) - \frac{1}{256\pi^2}
    \left[\log\frac{M_K^2}{\mu^2}+2\log\frac{M_\eta^2}{\mu^2}\right]
    \nonumber \\
&&\quad   -\frac{1}{256\pi^2(r-1)}
    \left(3\log\frac{M_\eta^2}{M_K^2}+5\log\frac{M_K^2}{M_\pi^2}\right)
\label{deltal5}\,.\\
\Delta L_6 
       &=& L_6^r(\mu) -\frac{1}{512\pi^2}\left(
       \log\frac{M_K^2}{\mu^2} 
       + \frac{2}{9}\log \frac{M_\eta^2}{\mu^2}
         \right) \nonumber \\
 && \quad    +\frac{1}{512 \pi^2}\frac{r}{(r+2)(r-1)} \left( 3 \log
          \frac{M_K^2}{M_\pi^2}  + \log \frac{M_\eta^2}{M_K^2} \right) \,.\label{deltal6}\\
 \Delta L_8 &=& L_8^r(\mu) - \frac{1}{512\pi^2}
    \left[\log\frac{M_K^2}{\mu^2}+\frac{2}{3}\log\frac{M_\eta^2}{\mu^2}\right]
 \nonumber \\
&& \quad    -\frac{1}{512 \pi^2 (r-1)} \left( 3 \log
          \frac{M_K^2}{M_\pi^2}  + \log \frac{M_\eta^2}{M_K^2} \right) \,, 
\label{deltal8}
\end{eqnarray}
The values of the logarithms are only mildly dependent on $r$; for $r=25$, 
\begin{eqnarray}
\Delta L_4=L_4^r(M_\rho)+0.51\cdot 10^{-3}\,, &\qquad&
\Delta L_5=L_5^r(M_\rho)+0.67\cdot 10^{-3}\,,\\
\Delta L_6=L_6^r(M_\rho)+0.26\cdot 10^{-3}\,, &\qquad&
\Delta L_8=L_8^r(M_\rho)+0.20\cdot 10^{-3}\,.
\end{eqnarray}

Since $F_\pi$, $F_K$, $M_\pi$ and $M_K$ are accurately known, we can use
these expressions to eliminate some of the $O(p^4)$ LECs in the chiral
expansion of other observables. This is rather
different from the usual $\chi$PT trading, since we keep 
explicitly higher-order terms that would have been neglected in the 
usual (perturbative) treatment of chiral series. 

From the masses and decay constants (\ref{fpi})-(\ref{fkamka}), 
we get the equivalent set of equations providing 
some  $O(p^4)$ LECs 
in terms of physical masses and decay constants, $r,X(3),Y(3)$ and NNLO 
remainders:
\begin{eqnarray} \label{idl6}
Y^2(3)\Delta L_6 &=& \frac{1}{16(r+2)}\frac{F_\pi^2}{M_\pi^2}
  [1-\epsilon(r)-X(3)-d]\,,
\\ 
\label{idl8}
Y^2(3)\Delta L_8 &=& \frac{1}{16}\frac{F_\pi^2}{M_\pi^2}
  [\epsilon(r)+d']\,,
\\ 
\label{idl4}
Y(3)\Delta L_4 &=& \frac{1}{8(r+2)}\frac{F_\pi^2}{M_\pi^2}
  [1-\eta(r)-Z(3)-e]\,,
\\ 
\label{idl5}
Y(3)\Delta L_5 &=& \frac{1}{8}\frac{F_\pi^2}{M_\pi^2}
  [\eta(r)+e']\,.
\end{eqnarray}
with
\begin{equation} \label{funcr}
\epsilon(r) = 2\frac{r_2-r}{r^2-1}, \qquad
\eta(r)=\frac{2}{r-1}\left(\frac{F_K^2}{F_\pi^2}-1\right)\,,
\end{equation}
and the following linear combinations of NNLO remainders arise:
\begin{eqnarray} \label{remainmass}
d  &=& \frac{r+1}{r-1}d_{\pi} - 
   \left(\epsilon(r)+\frac{2}{r-1}\right)d_K\,, \qquad d'=d-d_\pi\,,\\
e &=&\frac{r+1}{r-1}e_{\pi}-
 \left(\eta(r)+\frac{2}{r-1}\right)e_{K}\,,  \qquad e'=e-e_\pi\,.
\end{eqnarray}

The above identities are algebraically exact, but they are useful only as 
long as NNLO remainders are small. In refs.~\cite{resum,zr}, 
the size of the NNLO remainders was taken as
\begin{equation}
d,e = O(m_s^2) \sim 10\%\,, \qquad d',e' = O(mm_s) \sim 3\%\,,
\end{equation}
with the rule of thumb that NNLO corrections of size $O(m_s^2)$ should
not exceed $(30\%)^2\simeq 10\%$ of the contribution to the observable while
$O(m_s m)$ terms would be less than $30\% \cdot 10\% \simeq 3\%$. We will
propose in the next section a different but compatible way of dealing with this issue.

In eqs.~(\ref{idl6})-(\ref{idl5}), the presence of powers of 
$Y(3)$, i.e., $B_0$, follows from the normalisation of the scalar 
and pseudoscalar sources in
ref.~\cite{chpt3}: these powers arise only for $O(p^4)$ LECs related
to explicit chiral symmetry breaking (two powers for $L_6,L_7,L_8$,
one for $L_4$ and $L_5$), and are absent for LECs associated with
purely derivative terms.

\subsection{From bare expansions to Re$\chi$PT expansions}

As shown in detail in ref.~\cite{ordfluc}, plugging
eqs.~(\ref{idl6})-(\ref{idl5}) into the bare expansions for other observables
corresponds to resumming the vacuum fluctuations encoded in $L_4$ and
$L_6$. As an illustration, 
we recall that we can exploit  eqs.~(\ref{idl6})-(\ref{idl5}) 
to relate $Y(3)$ to the chiral couplings $L_4$ and $L_6$:
\begin{eqnarray}\label{eq:y3nonlin}
Y(3)&=&\frac{2[1-\epsilon(r)-d]}
  {\displaystyle[1-\eta(r)-e]+\sqrt{[1-\eta(r)-e]^2+ k\times
    [2\Delta L_6-\Delta L_4]}}\,,\\
k&=& 32(r+2)\frac{M_\pi^2}{F_\pi^2}[1-\epsilon(r)-d]\,.
\end{eqnarray}
If vacuum fluctuations are small, i.e. $\Delta L_6$ and $\Delta L_4$ 
almost vanishing, one can treat $k\times [2\Delta L_6-\Delta L_4]$ in the denominator 
as a small perturbation and linearise the equation as 
$Y(3)=1+ O(p^2)$.
This corresponds to the usual (iterative and perturbative) treatment of chiral series.
However the factor $k$ is very large 
($k\simeq 1900$ for $r=25$) and values of 
$\Delta L_6$ and $\Delta L_4$ of a few
$10^{-3}$ suffice to yield an important deviation of $Y(3)$ from 1, while
the linear approximation becomes inaccurate.
Similar relations exist between $X(3)$ and $\Delta L_6$, and between $Z(3)$ and
$\Delta L_4$~\cite{ordfluc}. 

Using eqs.~(\ref{idl6})-(\ref{idl5}), 
we  obtain the one-loop expansions of good observables in Re$\chi$PT, by
using eqs.~(\ref{idl6})-(\ref{idl5}) and
reexpressing $F(3)$, $mB_0$, $Y(3)L_4$, $Y(3)L_5$, $Y(3)^2 L_6$, $Y(3)^2 L_8$ 
(and $Y(3)^2 L_7$ through $\eta$ identities)
in terms of the three parameters of interest $X(3)$,
$Z(3)$, $r$ and NNLO remainders. In the case of $\pi\pi$ and $\pi K$
scatterings, only three $O(p^4)$ LECs ($L_1,L_2,L_3$) will remain.
The square root induced by equations like eq.~(\ref{eq:y3nonlin}) is a 
non-perturbative feature of our framework. It amounts to resumming 
(potentially) large contributions of vacuum fluctuations, 
encoded in the Zweig-rule violating LECs $L_4$ and $L_6$. This feature,
contrasting with the usual treatment of chiral series, has led us
to call our framework Resummed Chiral Perturbation Theory or Re$\chi$PT.

There is a price to pay for this extension of the chiral framework in the case
of large fluctuations of $s\bar{s}$ pairs, and 
the resulting competition between LO and NLO in the chiral counting: 
some usual $O(p^4)$ relations cannot be exploited anymore, 
because of our ignorance about their convergence.
For instance, the quark mass ratio $r=m_s/m$ ($m=m_u=m_d$) cannot be fixed 
from $M_K^2/M_\pi^2$ since we do not control the convergence
of its three-flavour chiral expansion. $r$ becomes a free parameter which
can vary in the range:
\begin{equation}
r_1= 2\frac{F_K M_K}{F_\pi M_\pi}-1 \sim 8 \leq r \leq 
r_2= 2\frac{F_K^2 M_K^2}{F_\pi^2 M_\pi^2}-1 \sim 36\,.
\end{equation}
Similarly, one cannot determine LECs or combinations of LECs through ratios
of observables. For instance, one should not use $F_K/F_\pi$ to
determine $L_5$ at $O(p^4)$, because we do not know if the chiral expansion of 
$F_K/F_\pi$ converges at all. Finally, the agreement of the pseudoscalar spectrum with the 
Gell-Mann--Okubo formula requires a fine tuning of $L_7$ (however, this fine tuning is also 
needed in the case of a dominant $N_f=3$ quark condensate 
and small vacuum fluctuations~\cite{resum}).

\section{$\pi\pi$ and $\pi K$ scattering amplitudes}

In this section, we are applying the Re$\chi$PT framework to two examples of 
Goldstone-boson scatterings : $\pi\pi$ scattering, 
which probes the structure of 
QCD vacuum in the $N_f=2$ chiral limit, and $\pi K$ scattering, 
which is linked with the $N_f=3$ chiral limit.

\subsection{One-loop expression in Re$\chi$PT} \label{sec:chiralseries}

In the isospin symmetry limit, the low-energy $\pi\pi$ scattering is 
described by a single Lorentz-invariant amplitude:
\begin{equation}
A(\pi^a(p_1)+\pi^b(p_2)\to\pi^c(p_3)+\pi^d(p_4))
 =\delta^{ab} \delta^{cd} A(s,t,u)
 +\delta^{ac} \delta^{bd} A(t,u,s)
 +\delta^{ad} \delta^{bc} A(u,t,s)\,
\end{equation}
where the usual Mandelstam variables are:
\begin{equation}
s=(p_1+p_2)^2\,, \qquad t=(p_1-p_3)^2\,, \qquad u=(p_1-p_4)^2\,,
\end{equation} 
and $A$ is symmetric under $t\leftrightarrow u$ exchange.
In a similar way, we consider the low-energy $\pi K$ scattering, which can be
decomposed into two amplitudes according to isospin in
the $s$-channel $I=3/2$ and $I=1/2$:
\begin{equation}
A(\pi^a(p_1)+K^i(p_2)\to\pi^b(p_3)+K^j(p_4))=F^I_{\pi K}(s,t,u)\,,
\end{equation}
from which one can define two amplitudes, respectively even and odd under 
$s\leftrightarrow u$ exchange:
\begin{eqnarray}
B(s,t,u)&=&\frac{2}{3}F^{3/2}_{\pi K}(s,t,u) 
                    +\frac{1}{3}F^{1/2}_{\pi K}(s,t,u)\,,\\
C(s,t,u)&=&-\frac{1}{3}F^{3/2}_{\pi K}(s,t,u) 
                    +\frac{1}{3}F^{1/2}_{\pi K}(s,t,u)\,.
\end{eqnarray}
In addition, crossing symmetry
provides a relation between the two amplitudes:
\begin{equation}
F^{1/2}_{\pi K}(s,t,u)=\frac{3}{2} F^{3/2}_{\pi K}(u,t,s)-\frac{1}{2} F^{3/2}_{\pi K}(s,t,u)\,.
\end{equation}

We can apply the prescriptions described in sec.~\ref{sec:rechipt} to determine
the one-loop Re$\chi$PT expansions of $A$, $B$ and 
$C$. The relevant good observables, which can be derived
from Green functions of vector/axial currents, are $F_\pi^4 A$,
$F_\pi F_K F$ and $F_\pi F_K G$. 

\begin{enumerate}
\item We determine the one-loop bare expansions of these quantities. 
This can be done using the generating functional of $N_f=3$ 
$\chi$PT~\cite{chpt3}, with
the essential difference that we keep the distinction between $O(p^2)$ 
truncated masses and physical masses of the Goldstone bosons. This was
performed in the case of $\pi K$ scattering in ref.~\cite{BKM}. A similar work
can be done in the case of $\pi\pi$ scattering. The corresponding (rather
lengthy) expressions are summarised in app.~\ref{app:bareexp}.

\item We use eqs.~(\ref{deltal4})-(\ref{deltal8}) to reexpress the 
$O(p^4)$ LECs $L_4$, $L_5$, $L_6$, $L_8$ in terms of $r$, $X(3)$ and $Z(3)$,
and NNLO remainders related to $\pi$ and $K$ masses and decay constants. 
We denote with the superscript ${LO+NLO}$ the resulting expressions, which include
the LO and NLO expansions of the relevant good observables and resum
the vacuum fluctuations encoded in $L_4$ and $L_6$. 

\item To obtain the Re$\chi$PT expansions of
the $\pi\pi$ and $\pi K$ scattering amplitudes,
we add to the resulting expressions a polynomial modeling higher-order
contributions : 
\begin{eqnarray} \label{eq:higha}
F_\pi^4 A^{Re\chi PT} &=& F_\pi^4 A^{LO+NLO}
   + F_\pi^2 (s_A-M_\pi^2) a_1
   + F_\pi^2 (s-s_A)a_2\\
&&\qquad   + (s-s_A)^2 a_3 + [(t-t_A)^2 + (u-u_A)^2] a_4\,, \nonumber\\
F_\pi^2 F_K^2 B^{Re\chi PT} &=&  \label{eq:highb}
   F_\pi^2 F_K^2 B^{LO+NLO}
   + F_\pi F_K t_B b_1
   + F_\pi F_K (t-t_B)b_2\\
&&\qquad      + (t-t_B)^2 b_3 
   + [(s-s_B)^2 + (u-u_B)^2] b_4\,, \nonumber\\
F_\pi^2 F_K^2 C^{Re\chi PT} &=&    \label{eq:highc}
   F_\pi^2 F_K^2 C^{LO+NLO}
   + F_\pi F_K (s-u) c_1
   + (t-t_B)(s-u) c_2 \,,
\end{eqnarray}
where $(s_A,t_A,u_A)$, $(s_B,t_B,u_B)$, $(s_C,t_C,u_C)$ denote the points
around which we perform the expansion of the NNLO polynomial. The first
remainder is multiplied by a constant estimating roughly
the value of the amplitude at the expansion point (obtained from the LO chiral
expression). The other remainders are multiplied by 
polynomials in the Mandelstam variables which vanish at the expansion point
and respect the crossing properties of the amplitude.
\end{enumerate}

For our purposes, we take:
\begin{eqnarray}
(s_A,t_A,u_A)&=&(4/3 M_\pi^2,4/3 M_\pi^2,4/3 M_\pi^2)\,, \\
(s_B,t_B,u_B)&=&(s_C,t_C,u_C)=(M_K^2+1/3 M_\pi^2,4/3 M_\pi^2,M_K^2+1/3 M_\pi^2)\,.
\end{eqnarray}

The remainders $a_i,b_i,c_i$ include only NNLO terms or higher : we expect therefore 
these contributions to be suppressed by $1/\Lambda_H^4$ where $\Lambda_H$ is a
typical hadronic scale~\cite{lecsize}. On the other hand, the numerator may depend on 
the remainder considered, but the contribution to the
polynomial must be order $O(p^6)$ in the usual chiral counting. This means that
the remainders have a typical size of order:
\begin{equation}
a_1,a_2,b_1,b_2,c_1 \sim \frac{M_K^4}{\Lambda_H^4}\,,
\qquad a_3,a_4,b_3,b_4,c_2 \sim \frac{F_\pi^2 M_K^2}{\Lambda_H^4}\,.
\end{equation}
Remainders associated with higher-order polynomials 
would be of order $F_\pi^4/\Lambda_H^4$,
much suppressed compared to the terms considered here, and thus
neglected in the following analysis.

In the case of $\pi\pi$ scattering, we can exploit the behaviour of the
amplitude in the $N_f=2$ chiral limit 
in order to constrain the size of NNLO remainders further. Indeed, from
$N_f=2$ chiral perturbation theory, we know that:
\begin{equation} \label{eq:nf2pipi}
F_\pi^4 A(s,t,u) - F_\pi^2 (s-M_\pi^2) = O(\epsilon^4) 
\quad {\rm with} \quad \epsilon^2 \sim p^2 \sim m\,.
\end{equation}
$\epsilon$ counts only powers of $m=m_u=m_d$ but not those of $m_s$. If we
compare this relation with $F_\pi^4 A$ expressed in $N_f=3$
Re$\chi$PT in eqs.~(\ref{eq:higha}) and (\ref{eq:rechiptpipi}), 
we see that the relation~(\ref{eq:nf2pipi}) implies a constrain
on the NNLO remainders : $a_1-e_\pi-(d_\pi-e_\pi)/3/(s_A/M_\pi^2-1)$
 and $a_2-e_\pi$ must be proportional to $m$. Therefore, we can 
expect the remainders to exhibit the typical sizes:
\begin{equation}
a_1-e-\frac{d-e}{3(s_A/M_\pi^2-1)},a_2-e \sim \frac{M_\pi^2 M_K^2}{\Lambda_H^4}\,, \qquad
b_1,b_2,c_1 \sim \frac{M_K^4}{\Lambda_H^4}\,,
\qquad a_3,a_4,b_3,b_4,c_2 \sim \frac{F_\pi^2 M_K^2}{\Lambda_H^4}\,.
\end{equation}

According to this discussion, we take the following ranges for the direct remainders:
\begin{eqnarray}
&& a_1-e-\frac{F_\pi^2M_\pi^2}{3(s_A-M_\pi^2)}(d-e),a_2-e \in 
        \left[-\frac{2 M_\pi^2 M_K^2}{\Lambda_H^4},\frac{2 M_\pi^2
        M_K^2}{\Lambda_H^4}\right]\,,
\label{eq:sigmadirect}
\\
&& b_1,b_2,c_1 \in \left[-\frac{M_K^4}{\Lambda_H^4},\frac{M_K^4}{\Lambda_H^4}\right]\,,
\qquad a_3,a_4,b_3,b_4,c_2  \in \left[-\frac{F_\pi^2
        M_K^2}{\Lambda_H^4},\frac{F_\pi^2 M_K^2}{\Lambda_H^4}\right]\,,
\nonumber
\end{eqnarray}
with $\Lambda_H=0.85$ GeV. This choice for the numerical value of $\Lambda_H$
provides a good agreement of our estimates with those
used in refs.~\cite{param,pipi,resum} for energy-independent
quantities. In the latter references, NNLO remainders were taken of order
$O(m_s^2)=(30\%)^2=10\%$  of the leading-order value, unless they were
suppressed by one power of $m$ and thus of order $O(mm_s)=30\%\times
10\%=3\%$. According to this work, the same remainders must remain respectively of order 
$M_K^4/\Lambda_H^4=12\%$ and  $2 M_\pi^2 M_K^2/\Lambda_H^4=2\%$. In addition,
one can check that the definition and size of remainders given in this section
can be applied to the two-point correlators related to $F_P^2$ and $F_\pi^2 M_\pi^2$
with an expansion around the point of vanishing transfer momentum, leading to
remainders identical to those defined in sec.~\ref{sec:massdec}.

\subsection{Roy and Roy-Steiner equations}

The above theoretical expressions for low-energy $\pi\pi$ and $\pi K$
scattering must be compared to experimental information in order to
extract the parameters of three-flavour chiral symmetry breaking. Fortunately,
dispersion relations provide an appropriate framework 
to analyse experimental data and extract the low-energy behaviour of the
amplitude, through Roy and Roy-Steiner equations.

In ref.~\cite{ACGL}, Roy equations were derived and solved with experimental
input on high-energy $\pi\pi$ scattering. The solutions were parametrised
in terms of two scattering lengths $a_0^0$ and $a_0^2$.
In refs.~\cite{ACGL} and \cite{pipi}, these solutions, and some of their
extensions, were exploited together with 
recent data on $\pi\pi$ scattering in order to determine
the low-energy structure of the amplitude with the best accuracy. 
Ref.~\cite{CGL} proposed to combine  $K_{\ell 4}$ data on $\delta_0^0-\delta_1^1$
supplemented with a theoretical constraint from the scalar radius of the pion.
This constraint was assessed critically in ref.~\cite{pipi}, where it
was proposed to avoid any reference to the scalar radius of the pion and to
rely only on experimental data, namely $K_{\ell 4}$ data supplemented with 
$I=2$ data. We follow the latter approach and take the results of the
so-called ``Global'' fit, eq.~(12) in ref.~\cite{pipi}, for
$\pi\pi$-scattering data.

In ref.~\cite{roypika}, Roy-Steiner equations were investigated to study the
$\pi K$ scattering amplitude. In spite of recent progress 
in $\tau\to K\pi\nu_\tau$ and $D\to K\pi e\nu_e$ decays, low-energy data
on $\pi K$ phase shifts is still lacking. However the dispersive
analysis of the data in the intermediate region turned out to provide rather tight
constrains on the low-energy $\pi K$ amplitude. We use the results of
ref.~\cite{roypika} for $\pi K$ scattering.

It is a straightforward, if tedious, exercise, to exploit the dispersive
representations of the amplitudes  $A$,$B$,$C$ found in sec.~3 of
ref.~\cite{ACGL} and in sec.~2 of ref.~\cite{roypika},
and to compute them in subthreshold regions, where none of 
the dispersion integrals exhibit singularities. We
checked in particular that our representation of the low-energy 
$\pi K$ amplitude was in good numerical agreement with the subthreshold expansion
presented in sec.~6.3 in ref.~\cite{roypika}.

We define the subthreshold region of interest for $\pi\pi$ scattering 
as a triangle in the Mandelstam plane delimited by points with $(s,t,u)$:
\begin{equation}
(2 M_\pi^2,M_\pi^2 , M_\pi^2 )\,, \qquad
(M_\pi^2/2,3/2 M_\pi^2,3/2 M_\pi^2)\,, \qquad
(M_\pi^2/2,3 M_\pi^2, M_\pi^2/2)\,,
\end{equation}
taking into account the symmetry of the amplitude under $t-u$ exchange. 
Similarly, we define for $\pi K$ scattering a
triangle in the Mandelstam plane with:
\begin{equation}
(M_K^2,2M_\pi^2,M_K^2) \,, \qquad
(M_K^2,0,M_K^2+2M_\pi^2) \,, \qquad
(M_K^2+M_\pi^2,0,M_K^2+M_\pi^2) \,,
\end{equation}
exploiting the symmetry or antisymmetry under $s-u$ exchange. In each triangle,
we defined 15 points regularly spaced where we compute the scattering
amplitudes. Some aspects of the computation, and of the correlations among the
points, are covered in app.~\ref{app:comp}.

\section{Matching in a frequentist approach}

We must match the chiral expansions of the scattering amplitudes
with the experimental values described in the previous section.
We perform this matching in a frequentist approach inspired
by the Rfit method~\cite{rfit}.

\subsection{Likelihood} \label{sec:likelihood}

We collect in a vector $V$ our $3n$ observables~:
\begin{equation} \label{eq:defv}
 V^T=\left[A(s_1,t_1),\ldots A(s_n,t_n),
         B(s'_1,t'_1),\ldots B(s'_n,t'_n),
         C(s''_1,t''_1),\ldots C(s''_n,t''_n)\right]\,.
\end{equation}

Since we use the masses and decay constant identities for pions and kaons
to reexpress the $O(p^4)$ LECs in terms of $F_P^2$ and $F_P^2M_P^2$
through eqs.~(\ref{idl6})-(\ref{idl5}), our
set of theoretical parameters is:
\begin{eqnarray}
{\rm Parameters} &:& r,X(3),Z(3),L_1^r,L_2^r,L_3\,,\\
{\rm Direct\ remainders} &:& a_1,a_2,a_3,a_4,b_1,b_2,b_3,b_4,c_1,c_2\,,\\
{\rm Indirect\ remainders} &:& d,d',e,e',d_X,d_Z\,.
\end{eqnarray}
We have separated the direct remainders, attached to the bare expansions of
the observables, and the indirect remainders, arising through the reexpression
of $O(p^4)$ LECs thanks to mass and decay constant equalities.
The latter include also the remainders $d_X$ and $d_Z$, whose expressions will
be given in sec.~\ref{sec:prior} and which are required to express 
the paramagnetic constraints on $X$ and $Z$, eq.~(\ref{eq:paramag}).

We construct the experimental likelihood ${\mathcal{L}}_{\rm exp}$, i.e. the probability of
observing the data for a given choice of theoretical 
parameters $T_n$ :
\begin{equation} \label{explike}
{\mathcal{L}}_{\rm exp}(T_n)=P({\rm data}|T_n)
   \propto \exp\left(-\frac{1}{2}(V_{th}-V_{exp})^T C^{-1} (V_{th}-V_{exp})\right)/\sqrt{\det C}\,.
\end{equation}
To avoid a proliferation of (purely numerical) normalisation factors of no
significance for our discussion, we use the sign $\propto$ meaning ``proportional to''.
$C$ is the covariance matrix between the experimental values $V_{exp}$
computed through eq.~(\ref{gencorrmat}), 
whereas $V_{th}$ denote the theoretical values computed with the particular
choice of $T_n$.
Since we expect strong correlations among the parameters, the covariance
matrix must be treated with some care, as described in app.~\ref{app:correl}.

The theoretical likelihood ${\mathcal{L}}_{\rm th}(T_n)$ describes our current knowledge on the theory
parameters. In agreement with the Rfit prescription~\cite{rfit}, we consider that 
${\mathcal{L}}_{\rm th}(T_n)=1$ if each theoretical parameter lies within its
allowed range described in the next section, otherwise the likelihood vanishes.

\subsection{Constraints on the theoretical parameters} \label{sec:prior}

To build the theoretical likelihood, we impose a list of constraints on the
theoretical parameters. Some constraints are fairly simple:
\begin{itemize}
\item We take the following range for the ratio of quark masses $r$:
\begin{equation}
r_1 \leq r \leq r_2\,, \qquad
  r_1=2\frac{F_KM_K}{F_\pi M_\pi}-1\,, \qquad
  r_2=2\left(\frac{F_KM_K}{F_\pi M_\pi}\right)^2-1\,.
\end{equation}

\item Vacuum stability yields constraints on the $N_f=3$ chiral order parameters:
\begin{equation}
X(3)\geq 0 \,, \qquad Z(3)\geq 0\,.
\end{equation}

\item We allow the three $O(p^4)$ LECs $L_1^r(M_\rho),L_2^r(M_\rho),L_3$ in
the range $[-F_\pi^2/\Lambda_H^2,F_\pi^2/\Lambda_H^2]$, i.e. 
lower than $12\cdot 10^{-3}$ in absolute value.

\item The direct remainders are constrained to remain in the range
given in eq.~(\ref{eq:sigmadirect}).

\item The indirect remainders must lie in the ranges discussed in
  sec.~\ref{sec:chiralseries}:
\begin{equation}
d',e',d_X,e_X  \in 
        \left[-\frac{2 M_\pi^2 M_K^2}{\Lambda_H^4},\frac{2 M_\pi^2
        M_K^2}{\Lambda_H^4}\right]\,,
\qquad
d,e \in \left[-\frac{M_K^4}{\Lambda_H^4},\frac{M_K^4}{\Lambda_H^4}\right]\,,
\end{equation}
i.e. 3\% for the first and 12\% for the latter.
\end{itemize}

A second set of constraints translates into bounds on combinations of remainders:
\begin{itemize}
\item Vacuum stability for $N_f=2$ chiral order parameters yields:
\begin{eqnarray}
X(2)\geq 0 &\leftrightarrow & d\leq d_{\rm max} \equiv 1-\epsilon(r)-Y(3)^2\times L_X \,,\\
Z(2)\geq 0 &\leftrightarrow & e\leq e_{\rm max} \equiv 1-\eta(r)-Y(3)\times L_Z \,,
\end{eqnarray}
where $L_X$ and $L_Z$ are small combinations of chiral logarithms denoted $f_1$ and $g_1$
in refs.~\cite{ordfluc,resum}. These chiral logarithms involve $M_K$ and
$M_\eta$ in the $N_f=2$ chiral limit, which can be computed through the
iterative method presented in ref.~\cite{ordfluc}.

\item The paramagnetic inequalities eq.~(\ref{eq:paramag}) lead to:
\begin{eqnarray}
X(3)\leq X(2) &\leftrightarrow & d_X \geq d_{X,{\rm min}} \equiv 1-\frac{d_{\rm max}-d}{X(3)(1-d)}\,,\\
Z(3)\leq Z(2) &\leftrightarrow & e_Z \geq e_{Z,{\rm min}} \equiv 1-\frac{e_{\rm max}-e}{Z(3)(1-e)}\,.
\end{eqnarray}

\item The ratio of order parameters $Y(3)=X(3)/Z(3)=2mB_0/M_\pi^2$ is bound~\cite{ordfluc}:
\begin{equation}
Y(3) \leq Y^{\rm max}=2\frac{1-\epsilon(r)-d}{1-\eta(r)-e}
\end{equation}

\end{itemize}

\subsection{Computation of the confidence level} \label{sec:cl}

Contrary to ref.~\cite{resum} which adopted a Bayesian approach to deal 
with $\pi\pi$ scattering, we follow the (frequentist) 
Rfit procedure advocated in ref.~\cite{rfit} and used for the analysis
of the CKM matrix in ref.~\cite{ckmfitter}.
From the theoretical and experimental likelihoods we define the function
of theoretical parameters
\begin{equation}
\chi^2(T_n)=-2\log {\mathcal{L}}(T_n) 
           =-2\log[{\mathcal{L}}_{\rm th}(T_n){\mathcal{L}}_{\rm exp}(T_n)]\,.
\end{equation}

We start by computing the absolute minimum of $\chi^2$, letting all
theoretical parameters vary freely : we denote $\chi^2_{{\rm min};{\rm all}}$ this
value. Then we focus on one particular theoretical parameter $T_i$. We
assume that it reaches a particular value $t_i$ and compute the minimum: 
\begin{equation}
\chi^2_{{\rm min};{\rm not\ }i}(t_i)=\min\{\chi^2(T_n);T_i=t_i\}\,.
\end{equation}
Then we compute the corresponding confidence level:
\begin{equation}
{\mathcal P}(t_i)={\rm Prob}[\chi^2_{{\rm min};{\rm not\ }i}(t_i)-\chi^2_{{\rm min};{\rm all}},1]\,,
\end{equation}
where ${\rm Prob}(c^2,N_{\rm dof})$ is the routine from the CERN library
providing the probability that a random variable having a 
$\chi^2$-distribution with $N_{\rm dof}$ 
degrees of freedom assumes a value which is larger than $c^2$.
Admittedly, we are simplifying the statistical problem at hand, since we
assume that the function $\chi^2(t_i)$ has indeed a $\chi^2$-distribution. This
should be a correct assumption if the experimental component is free from non
Gaussian contributions and inconsistent measurements~\cite{rfit}.

This method provides an upper bound on the marginal confidence level (CL)
of $T_i=t_i$ for the optimal set of theoretical parameters : 
the CL value is the probability that a new series of measurements
will agree with the most favourable set of theoretical
parameters (at $T_i=t_i$) in a worse way than the experimental results
actually used in the analysis~\cite{statlect}.
The value of $t_i$ for which ${\mathcal P}(t_i)$ is
maximal provides an estimator of $T_i$ : in the ideal case of very accurate
data in excellent agreement with theoretical expectations,
${\mathcal P}(t_i)$ should exhibit a sharp peak indicating the ``true'' value of $T_i$.

We have implemented this procedure in a program. Before turning to
Goldstone boson scattering, we checked the validity of our programs using ``fake'' observables. 
We designed observables with very simple chiral representations (linear or
quadratic dependence on $r,X(3),Z(3)$) 
and we simulated a set of data with a certain choice of $r,X(3),Z(3)$, adding some
random noise. 
We plugged these ``data'' into our program 
and computed the confidence level for each theoretical parameter $r,X(3),Z(3)$.
When the chiral representation of the observables depended on this parameter,
we obtained a function ${\mathcal P}(t_i)$ showing a peak in  
agreement with the value used to simulate the data (i.e., we recovered the
information contained in the data). When the chiral series for the observables
had no dependence on the parameter, the
function ${\mathcal P}(t_i)$ was flat (i.e., we did not extract information absent from the data). 

\section{Results} \label{sec:results}

In this section, we discuss the results obtained by matching the one-loop
Re$\chi$PT expansions and the dispersive results on $\pi\pi$ and $\pi K$
scattering, relying on the frequentist approach described in the previous
section.

\subsection{CL for order parameters and related quantities} \label{sec:rescls}

We have plotted the confidence level of the order
parameters $X(3)$, $Y(3)$ and $Z(3)$, as well as the quark mass ratio $r$. 
In each case, the dashed line indicates the results obtained
from $\pi\pi$ scattering, the dotted line from $\pi K$
  scattering, while the solid line stems from the combination of both pieces of information.

If we include $\pi \pi$ scattering only, we see that small values of $r$, below
13, are disfavoured (this is also the case for large values of $r$ above 25,
but not at a significant level) : $r\geq 12$ at 68~\% CL. 
The CL for $X(3)$ is flat up to 0.85, where it
suddenly drops, as well as that for $Z(3)$ up to 0.95. $Y(3)$, which is
related to $B_0$ and measures the fraction of the LO contribution to
$M_\pi^2$, is essentially not constrained, even though values close to
2 are slightly disfavoured.
If we consider $\pi K$ scattering only, $r$ and $Y(3)$ are essentially not constrained. Flat
CLs are observed for $X(3)$ and $Z(3)$, with a steep decrease respectively for 0.83
and 1. Finally, if we combine both pieces of information, intermediate values of $r$
are clearly favoured (between 20 and 25), in agreement with
the information contained in $\pi\pi$ and $\pi K$ scattering data. 
Low values of $X(3)$ and $Y(3)$ are preferred, 
whereas the CL for $Z(3)$ peaks around 0.8. We see that the combination of the
two data sets provides more stringent constraints on the various theoretical
parameters of interest (this issue is discussed in more detail in
app.~\ref{app:comb}), even though these results have still a limited 
statistical significance.

We recall that the frequentist method given here
provides an upper bound on the confidence level (CL) 
for the optimal set of theoretical parameters assuming $T_i=t_i$~\cite{rfit}.
In the ideal case, we would expect the CL to peak in a very limited interval
of $t_i$, providing the ``true'' value of the corresponding theoretical
parameter. In practice, we see that the chosen set of data is not
accurate enough to provide very stringent constraints on the theoretical
parameters. In such a case, the CL profiles can be exploited to extract
a confidence interval, say at 68~\% CL, i.e. a range of values so that the
probability that the range contains the true value of the parameter is
68~\%. This can be obtained by determining the region of parameter space 
where the CL curve lies above 0.32~\cite{statlect}. 

From the CL profiles obtained from the combined analysis of $\pi\pi$ and 
$\pi K$ scattering, we obtain the following confidence intervals at
68~\% CL :
\begin{equation}
r \geq 14.8\,, \qquad X(3)\leq 0.83\,, \qquad Y(3)\leq 1.1\,, 
  \qquad 0.18 \leq Z(3)\leq 1\,. \qquad  [68 \%\ {\rm CL}]
\end{equation}
The values for $L_1,L_2,L_3$ can also be determined in each case, and the 
corresponding confidence intervals are collected in table~\ref{table:Li}. 

\begin{table}
\begin{center}
\begin{tabular}{|c||c|c|c||c|c|c|}\hline
 & $\pi\pi$ data & $\pi K$ data & $\pi\pi$ and $\pi K$
   &Roy-Steiner $O(p^4)$ & $Kl_4,\ O(p^4)$ &  
$Kl_4,\ O(p^6)$ \\ \hline
$10^3\, L_1$  & $[-8.1,5.6]$ & $[-4.4, 4.1]$ & $[-2.1,2.2]$ 
      & $1.05\pm 0.12$ & $0.46\pm 0.24$ & $0.53\pm 0.25$ \\
$10^3\, L_2$  & $[0.2,2.4]$  & nd          & $[0,3.0]$    
      & $1.32\pm 0.03$ & $1.49\pm 0.23$ & $0.71\pm 0.27$ \\
$10^3\, L_3$  & nd           & nd          & $[-7.8,3.4]$ 
      & $-4.53\pm 0.14$ & $-3.18\pm 0.85$ & $-2.72\pm 1.12$\\
 \hline
\end{tabular}
\caption{\emph{Derivative chiral couplings $L_{1,2,3}^r(\mu)$ at $\mu=0.77$ GeV 
obtained in our approach. The confidence intervals correspond to a $68~\%$ 
CL. ``nd'' means that the corresponding CL is flat over the whole
range imposed by the theoretical likelihood, and thus the coupling is not
determined.
Also shown are results obtained assuming small vacuum fluctuations of 
$s\bar{s}$ pairs : ref.~\cite{roypika} analysed subthreshold 
$\pi K$ parameters from Roy-Steiner equations at order $p^4$ (col. 5), whereas
ref.~\cite{amoros} performed fits to the $Kl_4$ form factors
using chiral expansions at order $p^4$ (col. 6) 
as well as  $p^6$ (col. 7).\label{table:Li}}}
\end{center}
\end{table}

\begin{figure}
\begin{center}

\includegraphics[width=11cm]{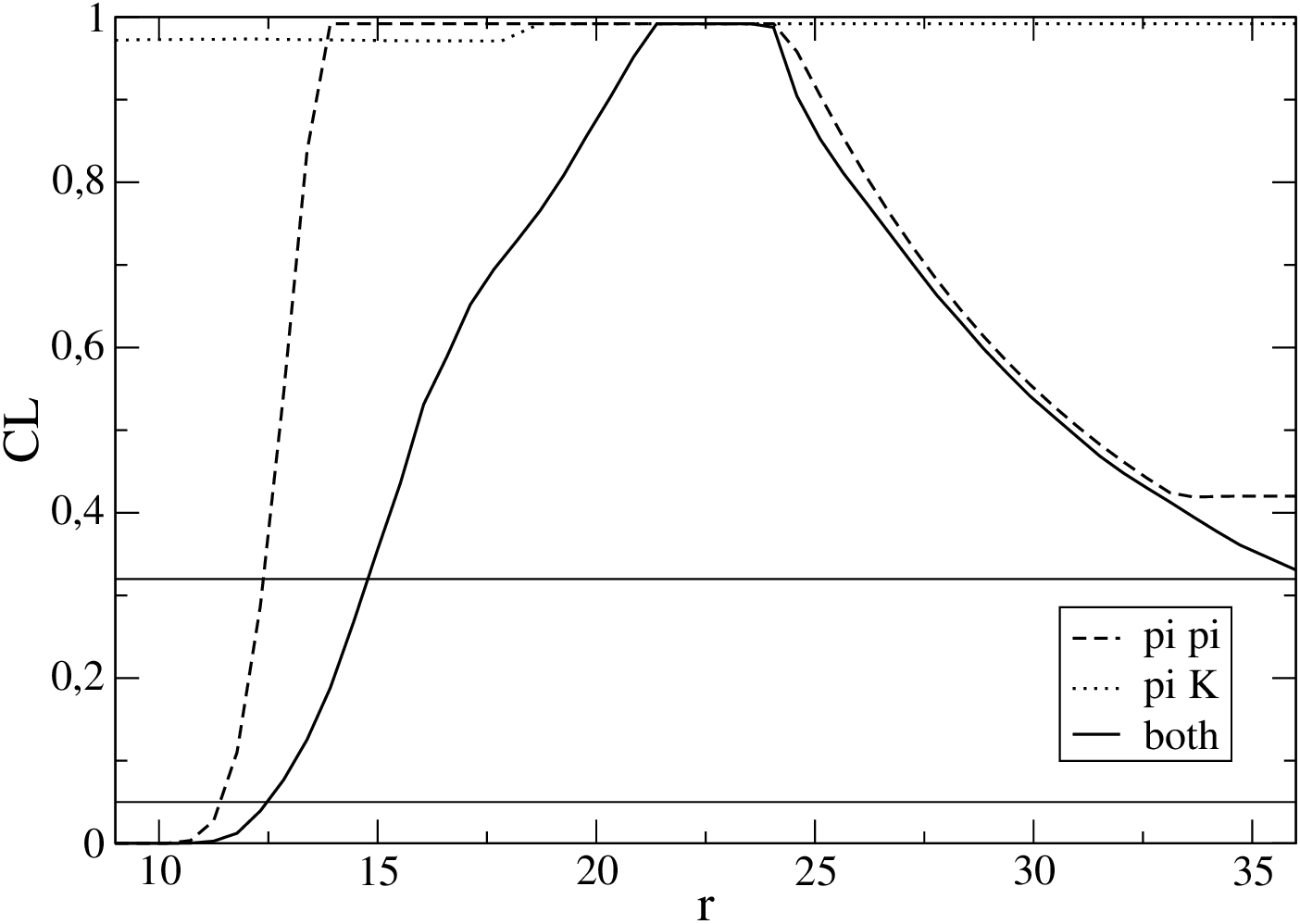}

\vspace{1.3cm}

\includegraphics[width=11cm]{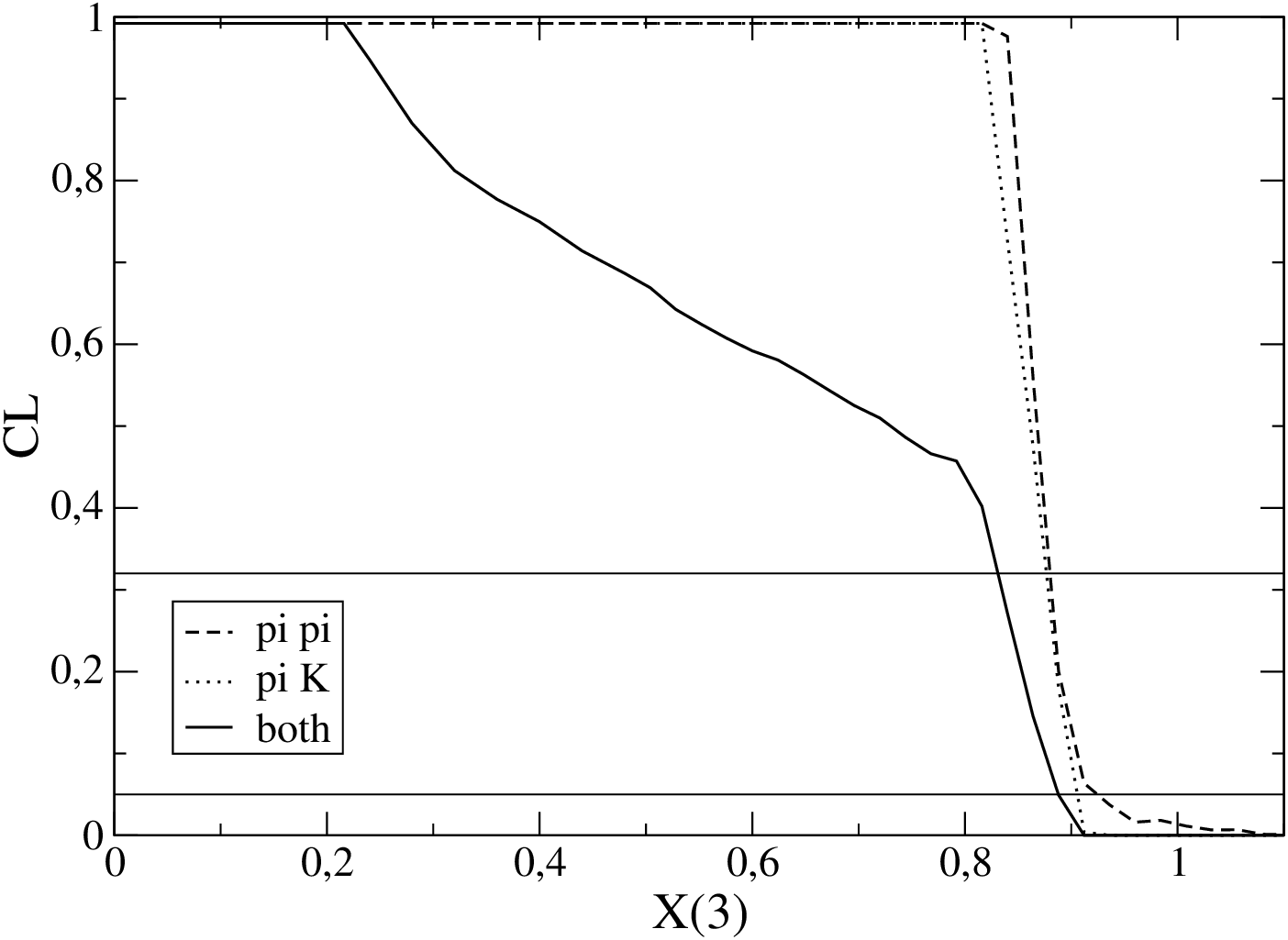}

\caption{CL profiles for $r=m_s/m$ (top) and $X(3)=2m\Sigma(3)/(F_\pi^2 M_\pi^2)$ (bottom). The dashed line corresponds
  to experimental information on $\pi\pi$ scattering, the dotted line to $\pi K$
  scattering, and the solid line to the combination of both sets. The two
  horizontal lines indicate the confidence intervals at 68 and 95\% CL.}
\label{fig:rx}
\end{center}
\end{figure}

\begin{figure}
\begin{center}

\includegraphics[width=11cm]{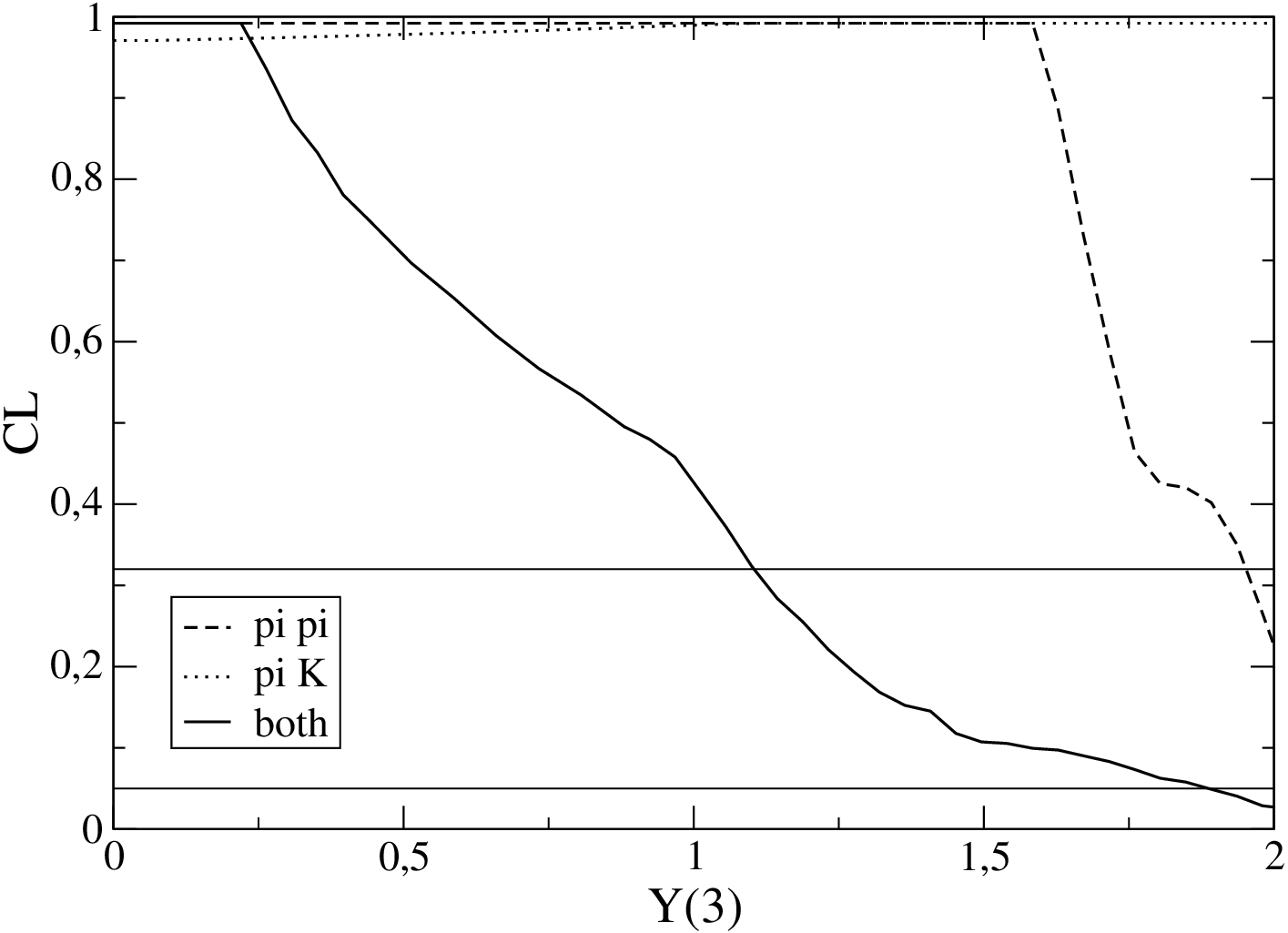}

\vspace{1.3cm}

\includegraphics[width=11cm]{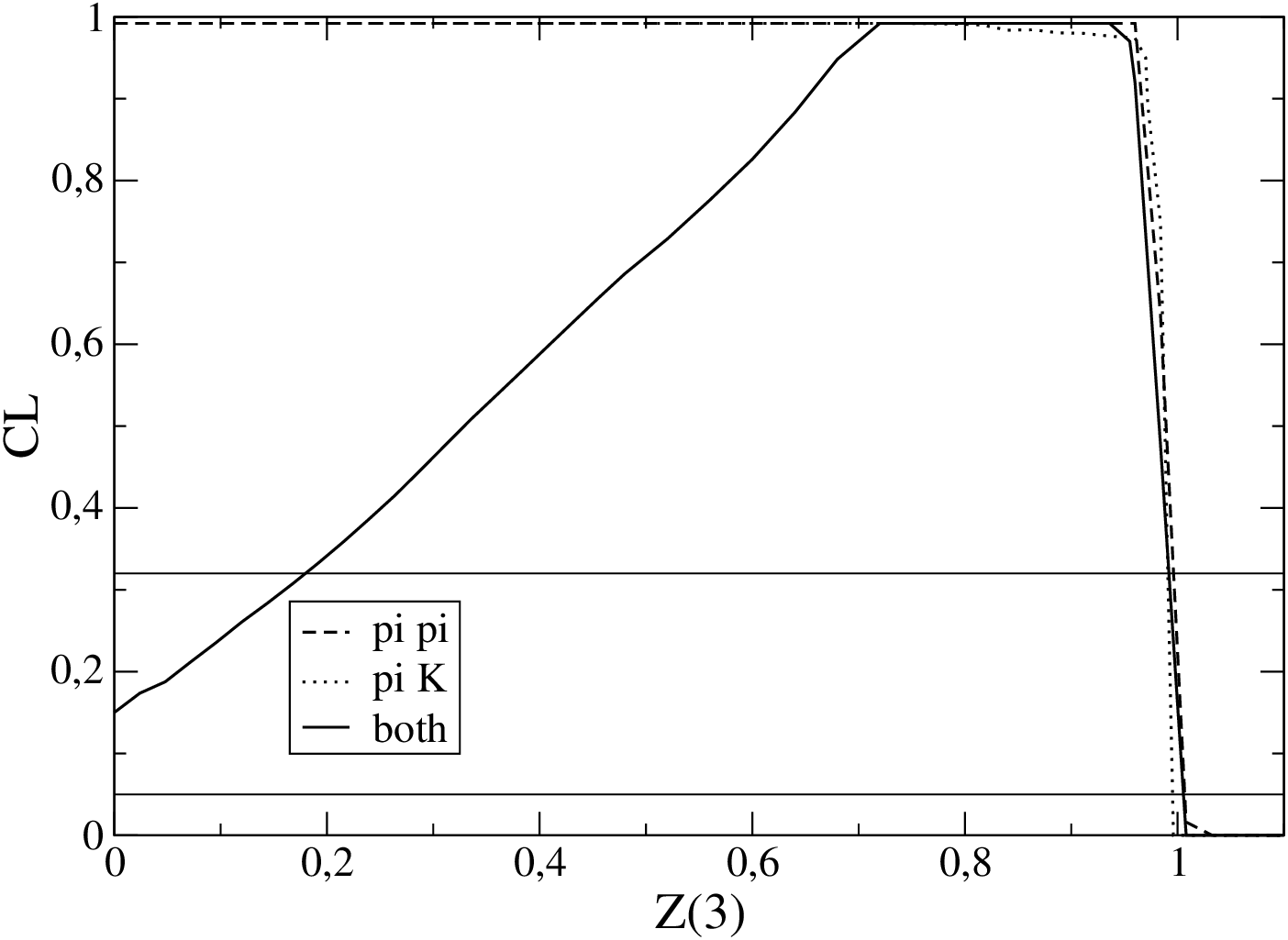}

\caption{CL profiles for $Y(3)=2mB_0/M_\pi^2$ (top) and $Z(3)=F_0^2/F_\pi^2$ (bottom). The dashed line corresponds
  to experimental information on $\pi\pi$ scattering, the dotted line to $\pi K$
  scattering, and the solid line to the combination of both sets. The two
  horizontal lines indicate the confidence intervals at 68 and 95\% CL.}

\label{fig:yz}
\end{center}
\end{figure}

As a cross-check, we have also studied the case where the higher-order direct
 remainders are removed, i.e. eqs.~(\ref{eq:higha})-(\ref{eq:highc}) is set to
 zero. The corresponding CLs are sharper, but very similar in shape to those
 presented here. Therefore, the polynomial terms modeling higher order contributions
 tend to push CLs towards 1, but the qualitative features
 shown in figs.~\ref{fig:rx} and \ref{fig:yz} stem mainly from the matching of
 LO and NLO terms of the Re$\chi$PT expansion to experimental information.

The scenario mildly favoured from the matching of both $\pi\pi$ and $\pi K$ scatterings would
correspond to a value of $r=m_s/m$ quite close to the canonical value
$r=25$. However, we emphasise that this agreement is rather coincidental : the
latter value comes from the (perturbative) reexpression of $M_K^2/M_\pi^2$ in
terms of $r$, assuming that the chiral expansions of the two squared 
masses converge quickly. 
This assumption is not supported by our results for the quark
condensate (or $X$), which exhibits some suppression when one moves
from the $N_f=2$ chiral limit to the $N_f=3$ one, 
i.e., when $m_s$ decreases from its physical value down to zero. 
On the other hand, the pion decay constant (or $Z$) seems quite
stable from $N_f=2$ to $N_f=3$, see eqs.~(\ref{eq:x2})-(\ref{eq:z2}). 
If our results are confirmed by further experimental data, we expect
the usual treatment of $N_f=3$ chiral expansions to yield
unstable expansions, with significant numerical competition among
terms of different orders in the chiral counting.

Such a situation is reminiscent of a scenario proposed some time ago concerning the
$N_f$-dependence of the chiral structure of QCD vacuum~\cite{param,nfdep}.
The quark condensate $\Sigma(N_f)$ and the decay constant $F(N_f)$ depend on the way small eigenvalues of the
Dirac operator accumulate around zero in the thermodynamic limit. It was conjectured that the two order
parameters could decrease at a different rate when the number of massless
flavours $N_f$ increases : the quark condensate would vanish first,
followed later by the vanishing of the decay constant related to the
restoration of chiral symmetry.
The trend of our results for $N_f=3$ order parameters,
compared to $N_f=2$ results, could fit such a scenario, but more data
should be included in the analysis before we reach statistically significant
CLs for the various theoretical parameters analysed here.

\subsection{Comparison with some earlier works}

\subsubsection{$\pi\pi$ scattering}

For $\pi\pi$ scattering, it 
is interesting to compare our results with ref.~\cite{resum}, which shares 
some ideas and issues with the present paper. This work
differs on three points from ref.~\cite{resum} : we include $\pi K$ 
scattering in our analysis, we choose as observables
the scattering amplitudes in subthreshold regions rather than the subtraction
constants involved in dispersive representations, we perform the statistical
analysis in a frequentist framework rather than a Bayesian one.

We observe the same qualitative
features in both analyses. 
As expected, low values of $r$ are strongly disfavoured. Indeed, the analysis of
currently available data on $\pi\pi$ scattering~\cite{pipi} provides a value
of $X(2)$, eq.~(\ref{eq:x2}). As illustrated in fig.~1 of ref.~\cite{param},
$X(2)$ is related to $r$ through the pion and kaon mass and decay constant identities,
eqs.~(\ref{fpi})-(\ref{fkamka}): the value of $X(2)$ from ref.~\cite{pipi} 
favours the same range for the quark mass ratio as the upper plot in
fig.~\ref{fig:rx}. On the other hand, we find that $X(3)$ and $Z(3)$ are only constrained
through an upper bound, in numerical agreement with the paramagnetic 
inequalities $X(3)\leq X(2)$ and $Z(3)\leq Z(2)$.

This agreement is particularly gratifying since the method of analysis of the
present work does not require computing any $N_f=2$ chiral order parameters or related 
subtraction constants like refs.~\cite{pipi,resum}. Moreover, one can see an
improvement compared to the latter references, thanks to the frequentist
approach chosen here.  In
ref.~\cite{resum}, it was difficult to disentangle the effect of the data
from that of the Bayesian priors inside a posterior p.d.f.~: 
the so-called ``reference profiles''
(p.d.f.s from priors but no data) had to be compared to
the posterior p.d.f.s (p.d.f.s from priors and data) to judge the impact
of $\pi\pi$ data. In the present paper, this intricate procedure and the
arbitrariness induced by Bayesian priors are avoided~:
it is clearly seen that
$\pi\pi$ data constrains $X(3)$ and $Z(3)$ only through the values of $X(2)$
and $Z(2)$ and the corresponding paramagnetic upper bounds.

\subsubsection{$\pi K$ scattering}

For $\pi K$ scattering, we can compare our results with
ref.~\cite{roypika}, where the solutions of the Roy-Steiner dispersion
relations were used to reconstruct the amplitudes in the subthreshold
region. These amplitudes were expanded around the point $s=u$, $t=0$, and
the coefficients of the polynomials, $C^+_{ij}$ and $C^-_{ij}$, were 
matched with their NLO chiral expansions in order to determine some $O(p^4)$ 
LECs. This led to a determination of $L_1,L_2,L_3$ recalled in the previous
section, and to a value of $L_4$ suggesting a significant suppression of
$Z(3)$. The value of $L_6$, though affected by large uncertainties, indicated
also a suppression of $X(3)$, stronger than that of $Z(3)$ :
\begin{equation}
{\rm Ref.}~\cite{roypika}: L_4^r(M_\rho)= (0.53\pm 0.39) \cdot 10^{-3}\,,
  \qquad [2L_6^r+L_8^r](M_\rho)=(3.66 \pm 1.52)\cdot 10^{-3}\,.
\end{equation}
Using eq.~(\ref{eq:y3nonlin}) and the other results of sec.~3.1 in
ref.~\cite{resum}, and taking 
$L_8^r(M_\rho)=(0.9\pm 0.3)\cdot 10^{-3}$~\cite{chpt3}
we can convert these results into
the parameters of interest: following these results, $X(3)$ 
would stand between 0.15 and 0.41, $Z(3)$ between 0.14 and 0.92, 
and $Y(3)$ between 0.44 and 1.05. Obviously, the low values of $X(3)$ and
$Z(3)$ indicate that the values obtained in ref.~\cite{roypika}, relying
on the assumption of small vacuum fluctuations and on $X(3)$ and $Z(3)$ close
to 1, should be reassessed relaxing this hypothesis.

If our results for the combined $\pi\pi$ and $\pi K$ data point towards a
similar pattern, our analysis of $\pi K$ data alone provides weaker
constraints than that of ref.~\cite{roypika}. At least two different reasons
lead us to weaker constraints. First, we have explicitly take into account the
presence of NNLO contributions which were neglected in the $O(p^4)$ analysis
of ref.~\cite{roypika} and which may affect significantly the energy-dependent
part of the amplitudes. Secondly, the analysis in ref.~\cite{roypika} assumes
explicitly the smallness of vacuum fluctuations : once we drop this
assumption, a smaller value of $L_4$ (and thus a value of $Z(3)$ close to 1)
can be compensated by the variation of other parameters, such as the quark
mass ratio $r$. These two phenomena may explain the weaker constraints
observed in our analysis.

\subsubsection{Combined analyses}

For the combined analysis of $\pi\pi$ and $\pi K$ scatterings, we can
compare our results with refs.~\cite{twolooppipi,twolooppika}. The
authors took a different approach from ours,
computing NNLO chiral expansions to $\pi\pi$ and $\pi K$ scattering amplitudes,
and matching with results on $\pi\pi$ scattering (scattering lengths) and
$\pi K$ scattering (scattering lengths and subthreshold expansion
coefficients), supplemented with information on $K_{\ell 4}$ form factors. 
In agreement with the one-loop framework of ref.~\cite{chpt3},
these two-loop computations assume a numerical dominance of LO contributions
and a quick convergence of $N_f=3$ chiral expansions. 

In previous studies in this NNLO framework~\cite{twolooprev}, the authors
performed fits to pseudoscalar masses and decay constants~\cite{twoloopmass}, 
$K_{\ell 3}$ decays~\cite{twoloopkl3}, and scalar form
factors~\cite{twoloopscal}. In each case, the values of the Zweig-rule
suppressed $O(p^4)$ LECs $L_4$ and $L_6$ had to be fixed by hand : fits of
similarity quality could be obtained with values of these two constants
corresponding either to small or large vacuum fluctuations of $s\bar{s}$
pairs. For scalar form factors, values of $L_4$ and $L_6$ larger
than conventionally assumed led to an improvement in the convergence of
observables (fits A,B,C compared to fit 10, in Table 2 of ref.~\cite{twoloopscal}).

In the case of refs.~\cite{twolooppipi,twolooppika}, the authors analysed
$\pi\pi$ and $\pi K$ scattering amplitudes in the same NNLO framework.
The fits were not able to reproduce some observables, in 
particular among $\pi K$ subthreshold coefficients. A particular subset of 
subthreshold coefficients and scattering lengths led 
to $L_4^r(M_\rho)\simeq 0.2 \cdot 10^{-3}$ and $L_6(M_\rho)$ negative. Such
values correspond to $F_0$ rather small compared to $F_\pi$, with a rather 
unsatisfying convergence of some observables : for instance, the pion
mass exhibits instabilities in its chiral expansion~\cite{twolooprev}. 
It proves difficult to draw a fully consistent
picture for the structure of QCD vacuum in the $N_f=3$ chiral limit from these
results.

Some of the problems encountered in refs.~\cite{twolooppipi,twolooppika}
were reassessed in
ref.~\cite{naturalp6}, in particular the determination of NNLO LECs. 
Following ref.~\cite{satur}, the many $O(p^6)$ LECs are
often estimated using resonance saturation. 
In ref.~\cite{naturalp6}, the specific resonance Lagrangian used in
refs.~\cite{twolooppipi,twolooppika} was shown to provide values for
vector-dominated 
LECs rather far away from the expectations based on $\pi K$
dispersion relations, but other resonance Lagrangians failed also to reproduce
these same results. Therefore, one may wonder whether the problems of
convergence seen in \cite{twolooprev} could stem from two different sources.
The first one consists in 
the use of resonance saturation to fix $O(p^6)$ counterterms, which is
already delicate in vector channels and certainly questionable in the scalar
sector. The second one is the observed slow convergence of chiral
expansions, which contradicts the starting assumptions of the NNLO analysis. 
A comparison of Re$\chi$PT expansions with the NNLO formulae in 
refs.~\cite{twolooppipi,twolooppika} should highlight how large values
of the $O(p^4)$ LECs $L_4$ and $L_6$ might destabilise NNLO expansions and how
the explicit resummation of vacuum fluctuations of our work echoes in the
perturbative expansion adopted in the latter references.

\subsubsection{Lattice}

Other interesting developments are awaited from lattice simulations. The
effects presented in this paper are related to strange sea-quarks, and can
be tackled only with (2+1) dynamical fermions with light
masses. Unfortunately, fermions with interesting
chiral properties 
(Wilson, Ginsparg-Wilson, twisted-mass)~\cite{lattice} are still 
with at most two dynamical flavours. On the other hand, staggered
fermions~\cite{MILC} have been exploited for simulations with (2+1) dynamical
quarks, but their use is under much debate~\cite{staggered}. The presence of the
fourth root of the fermion determinant yields non-localities which are not
understood yet : at best, recovering QCD requires taking the various continuum 
limits in a very careful way.

A staggered version of 
chiral perturbation theory~\cite{SChPT} has been developed to
extract chiral LECs from the pseudoscalar spectrum. It attempts at reproducing
the fourth-rooting of the fermion determinant and includes many other effects
(lattice spacing, finite-volume effects, taste-breaking terms), leading
to a number of LECs much larger than in continuum unstaggered $\chi$PT. The
hope is that the LECs common to both theories should be identical because
QCD ought to be recovered as a limit of lattice QCD with fourth-rooted staggered fermions.
In practice~\cite{MILC}, chiral fits to staggered data on the pseudoscalar spectrum
must include a large number of parameters and thus are highly
non-trivial. Mixed actions with domain-wall valence quarks and staggered sea
quarks have also been considered to reduce the number of LECs involved
in the associated chiral Lagrangian at the price of losing unitarity 
in addition to locality~\cite{mixed}.

Bearing all these remarks in mind, we 
can focus on the following staggered values :
\begin{equation}
{\rm Ref.}~\cite{MILC}: \quad
r=27.2(4)\,,\quad [2L_6^r-L_4^r](M_\eta)=0.5(1)(2)\cdot 10^{-3}\,, \quad 
L_4^r(M_\eta)=0.1(2)(2)\cdot 10^{-3}\,.
\end{equation} 
Combining the errors in quadrature and using 
eq.~(\ref{eq:y3nonlin}) and the other results of sec.~3.1 in
ref.~\cite{resum}, we can convert these results into
the parameters of interest: following these lattice results, $X(3)$ 
would stand between 0.55 and 0.95, $Z(3)$ between 0.57 and 1.04, 
and $Y(3)$ between 0.67 and 1.08, values which are not in
striking disagreement with our results. Obviously, if the values of $X(3)$ and
$Z(3)$ are on the smaller end of these ranges, i.e., if $L_6$ and $L_4$ are
in the upper end of the range in ref.~\cite{MILC}, the assumption of
small vacuum fluctuations is not correct, and the extraction of the LECs 
by the means of staggered $\chi$PT should be reassessed more carefully.

As an alternative to such tests, which rely strongly on the usual treatment
of chiral series, we proposed a lattice test of the size of $s\bar{s}$ vacuum 
fluctuations based on Re$\chi$PT in ref.~\cite{lattest}. We considered
simulations with (2+1) flavours, with a strange quark mass at its physical
value, but two $u,d$ light quarks with identical masses $\tilde{m}$ 
larger than their physical values $m$ and smaller than $m_s$. 
The larger values of the $u,d$ masses enhanced the impact 
of the vacuum fluctuations encoded in $L_4$ and $L_6$ on
observables such as the masses and decay constants of pions and kaons. This
led to a difference in the curvatures of $F_P^2$ and $F_P^2 M_P^2$ ($P=\pi,K$)
as functions of $q=\tilde{m}/m_s$, depending on the size of $X(3)$ and $Z(3)$. 
The effect was less pronounced in the case of $M_P^2$, obtained as the ratio
of the two former observables, leading to a fairly linear behaviour as
a function of $q$. 

We proposed in the same reference a test of the size of
$X(3)$ on the lattice from the pion and kaon spectrum, by considering
the dependence on $q$ of the ratios :
\begin{equation}
R_\pi=\frac{\tilde{F}_\pi^2 \tilde{M}_\pi^2}{q F_\pi^2 M_\pi^2}
\qquad R_K=\frac{2\tilde{F}_K^2 \tilde{M}_K^2}{(q+1) F_K^2 M_K^2}
\end{equation}
where $\tilde{F}_\pi^2$ and $\tilde{M}_\pi^2$ denote quantities computed on
the lattice with $u,d$ quarks of mass $\tilde{m}$. We assessed the leading
finite-volume effects to conclude that large volumes (of side around 2.5 fm)
were required to tame these effects.

In any case, 
more dedicated studies on (2+1) fermions with different actions, lattice 
spacings and volumes will be required in order to draw definite 
conclusions from lattice simulations on the structure of $N_f=3$ chiral
vacuum. 

\section{Conclusion}

Vacuum fluctuations of $s\bar{s}$ pairs can induce significant differences
in the pattern of chiral symmetry breaking between the two conceivable chiral limits:
$N_f=2$ ($m_u=m_d=0$ but $m_s$ kept at its physical values) and $N_f=3$ 
($m_u=m_d=m_s=0$). These fluctuations might lead to a paramagnetic suppression of the two main chiral order parameters
in the $N_f=3$ chiral limit, the quark condensate and the pseudoscalar
decay constant, compared to their 
$N_f=2$ counterparts~\cite{param,ordfluc}.
Then, we would observe a numerical competition between
leading-order (LO) and next-to-leading order (NLO) contributions in chiral series, 
through the two $O(p^4)$ LECs $L_4$ and $L_6$ related to the violation of the Zweig
rule in the scalar sector.

In order to shed light on the size of these fluctuations, we developed
and modified the framework sketched in ref.~\cite{resum} : Resummed Chiral
Perturbation Theory or Re$\chi$PT. We applied it to 
our current knowledge of low-energy $\pi\pi$ and $\pi K$ scatterings.
First, we recalled and detailed our
treatment of one-loop chiral series in the case of large vacuum fluctuations: only
a subset of ``good'' observables is assumed to converge globally 
(so that NNLO contributions are much smaller than the sum of LO and NLO
contributions),
the chiral series must be 
treated in a particular way to derive bare expansions an resum the effects
of vacuum fluctuations, while NNLO remainders are introduced to keep track of
higher-order contributions. Then, in this resummed framework, called Re$\chi$PT, we
determined the one-loop expansions for $\pi\pi$ and $\pi K$ scattering amplitudes.
Relying on our current experimental knowledge,
we exploited solutions of Roy and Roy-Steiner equations within dispersive
representations to determine the values of the amplitudes in subthreshold
(unphysical) regions where chiral expansions should converge.

The two representations of the scattering amplitudes were matched in a
frequentist approach (inspired by Rfit~\cite{rfit}). The output of this
analysis are marginal CL curves, 
providing an upper bound on the confidence level (CL) 
for the optimal set of theoretical parameters at fixed $T_i=t_i$: 
the CL value is the probability that a new series of measurements
will agree with the most favourable set of theoretical
parameters (at $T_i=t_i$) in a worse way than the experimental results
actually used in the analysis~\cite{statlect}.

Unfortunately, the marginal CL profiles do not provide sharp peaks and thus
stringent constraints on theoretical parameters at a statistically 
significant level. However, our results point towards
some favoured regions of parameter space, see figs.~\ref{fig:rx} and
\ref{fig:yz}. If only~$\pi\pi$ scattering is included, 
the results obtained in earlier works~\cite{resum} are recovered : small
values of $r$ are disfavoured, whereas $X(3)$ and $Z(3)$ are only constrained
to remain below their $N_f=2$ counterpart due to paramagnetic
inequalities eq.~(\ref{eq:paramag}). 
$\pi K$ scattering alone does not constrain strongly the various
theoretical parameters, apart from setting bounds on $X(3)$ and $Z(3)$. The
combination of the two pieces of information proves more interesting :
the CL profile for $r$ peaks around 23, low values of $X(3)$ are preferred,
whereas the CL for $Z(3)$ exhibits a broad peak around 0.8. 

From the CL curves obtained from the combined analysis of $\pi\pi$ and 
$\pi K$ scattering, we obtain the following confidence intervals at
68~\% CL :
\begin{equation}
r \geq 14.8\,, \qquad X(3)\leq 0.83\,, \qquad Y(3)\leq 1.1\,, 
  \qquad 0.18 \leq Z(3)\leq 1\,. \qquad  [68 \%\ {\rm CL}]
\end{equation}
corresponding to the regions of parameter space where the marginal 
CL profiles lie above 0.32~\cite{statlect}.

The pattern of the marginal CL profiles is consistent with the scenario of 
significant vacuum fluctuations of $s\bar{s}$ pairs.
It reminds one of the interesting possibility that the decrease of 
order parameters from $N_f=2$ massless flavours to
$N_f=3$ is steeper in the case of the pseudoscalar decay constant
$F^2(N_f)$ than for the quark condensate $\Sigma(N_f)$.

The present analysis constitutes a first attempt to
analyse data with a limited statistical significance, and
it relies strongly on the experimental results gathered
on $\pi\pi$ and $\pi K$ scatterings. For $\pi\pi$ scattering, new results are
expected from the NA48 collaboration on $K_{\ell 4}$ decays~\cite{NA48} and
on the cusp in $K\to 3\pi$~\cite{expercusp,cusp}. For $\pi K$ scattering,
we hope to obtain more precise information from $D_{\ell 4}$
decays~\cite{experdl4,dl4} and $\tau\to K\pi\nu_\tau$
decays~\cite{expertau,tau}. In addition,
lattice studies could soon provide results for three
light flavours with well-controlled actions in the chiral regime.
These new high-accuracy data should shed
some more light on the chiral structure of QCD vacuum, and in particular on
its dependence on the number of massless flavours and 
the role played by the vacuum fluctuations of $s\bar{s}$ pairs.

\section*{Acknowledgments}
It is a pleasure to thank L.~Girlanda, J.J.~Sanz-Cillero and 
J.~Stern for collaboration in the early stages of this work, 
D.~Becirevic and B.~Moussallam 
for fruitful discussions and N.H.~Fuchs for many useful suggestions on the
manuscript.
This work was supported in part by the EU Contract No. MRTN-CT-2006-035482, \lq\lq FLAVIAnet''.

\appendix

\section{One-loop bare expansions of scattering amplitudes}
\label{app:bareexp}

\subsection{$\pi\pi$ scattering amplitude}

Following the prescription in sec.~\ref{sec:rechipt}
we obtain, for instance from ref.~\cite{KMSF} :
\begin{eqnarray}
F_\pi^4 A_{\pi\pi} &=& 
\frac{2}{3} mB_0F_0^2
+ F_0^2 \left( s - \frac{4}{3} M_{\pi}^2\right) \\ 
&& + \mu_\pi F_0^2 \left[ - 4 \left( s - \frac{4}{3} M_\pi^2 \right) -
  2 B_0 m \right] 
  + \mu_K F_0^2 \left[ - 2 \left( s - \frac{4}{3} M_\pi^2 \right) -
  \frac{4}{3} B_0 m \right] \nonumber \\
&& -\frac{2}{9} \mu_\eta F_0^2 B_0 m \nonumber \\
&& +16 B_0 m L_4^r \left[ \left(s - \frac{4}{3}M_\pi^2 \right) (r+4) -
  \frac{4}{3} M_\pi^2 \right]  + 32 B_0 m L_5^r \left( s - \frac{5}{3}
M_\pi^2 \right) \nonumber \\
&& +\frac{64}{3} B_0^2 m^2 L_6^r ( r + 8) + \frac{256}{3} B_0^2 m^2 L_8^r
\nonumber \\
&& + 4 \left( 2 L_1^r + L_3^r \right) \left(s - 2 M_\pi^2 \right)^2 + 4
L_2^r \left[ (t - 2 M_\pi^2)^2 + ( u - 2 M_\pi^2)^2\right] \nonumber \\
&& + \frac{1}{2} \left[ (s - 2 M_\pi^2)^2 + 8 B_0 m (s-2 M_\pi^2) +12
  B_0^2 m^2 \right] J^r_{\pi\pi}(s) \nonumber \\
&& +\frac{1}{4} \left[ (t - 2 M_\pi^2)^2 J^r_{\pi\pi}(t) +(u-2
  M_\pi^2)^2 J^r_{\pi\pi}(u) \right] \nonumber \\
&& + \frac{1}{8} \left[ (s - 2 M_\pi^2)^2 + 8 B_0 m (s-2 M_\pi^2) +16
  B_0^2 m^2 \right] J^r_{KK}(s) \nonumber \\
&& + \frac{2}{9} B_0^2 m^2 J^r_{\eta\eta}(s) \nonumber \\
&& + \frac{1}{2} \left[ (s-u) t \left( 2 M^r_{\pi\pi} +
  M^r_{KK}\right) (t) + (s-t) u \left( 2 M^r_{\pi\pi} +
  M^r_{KK}\right) (u) \right]\nonumber
\end{eqnarray}
where $\mo_P^2$ denotes the leading-order pseudoscalar squared mass 
of the Goldstone boson $P$ and the tadpole logarithm is
\begin{equation}
\mu_P=\frac{\mo_P^2}{32 \pi^2 F_0^2} \log \frac{M_P^2}{\mu^2}
\end{equation}
We recast the amplitude in the following form
\begin{eqnarray} \label{eq:rechiptpipi}
F_\pi^4 A_{\pi\pi} &=&
  \mathcal{A} + \left(s-\frac{4}{3} M_\pi^2\right) \times \mathcal{B}\\
&& + 4(2L_1^r+L_3) (s-2M_\pi^2)^2 +
  4L_2^r[(t-2M_\pi^2)^2+(u-2M_\pi^2)^2] \nonumber\\
&& +
  \frac{1}{2}[(s-2M_\pi^2)^2+8mB_0(s-2M_\pi^2)+12m^2B_0^2]J^r_{\pi\pi}(s)
\nonumber\\
&& +\frac{1}{4}[(t-2M_\pi^2)^2 J^r_{\pi\pi}(t)
               +(u-2M_\pi^2)^2 J^r_{\pi\pi}(u)]\nonumber\\
&& +\frac{1}{8}
      [(s-2M_\pi^2)^2+8mB_0(s-2M_\pi^2)+16m^2B_0^2]J^r_{KK}(s)\nonumber\\
&& +\frac{1}{18} 4m^2B_0^2 J^r_{\eta\eta}(s)\nonumber\\
&& +\frac{1}{2}
    [(s-u)t\times(2M^r_{\pi\pi}+M^r_{KK})(t)
     +(s-t)u\times(2M^r_{\pi\pi}+M^r_{KK})(u)]\nonumber
\end{eqnarray}
where $\mathcal{A}$ and $\mathcal{B}$ are scale-dependent
combinations of LECs :
\begin{eqnarray}
3\times {\mathcal{A}} &=& 2mB_0F_0^2 + 64m^2B_0^2[(r+8)L_6^r+4L_8^r]
  -32mB_0M_\pi^2[2L_4^r+L_5^r]\\
&&  -\frac{1}{32\pi^2}4m^2B_0^2 \left[
    3\log\frac{M_\pi^2}{\mu^2}+(r+1)\log\frac{M_K^2}{\mu^2}
    +\frac{1}{9}(2r+1)\log\frac{M_\eta^2}{\mu^2}
    \right]\nonumber\\
\mathcal{B} &=& F_0^2+16mB_0[(r+4)L_4^r+2L_5^r]\\
&&  -\frac{1}{32\pi^2}2mB_0 \left[
    4\log\frac{M_\pi^2}{\mu^2}+(r+1)\log\frac{M_K^2}{\mu^2}
    \right]\nonumber
\end{eqnarray}
which correspond to $F_\pi^2 M_\pi^2\alpha^r_{\pi\pi}/3$ 
and $\mathcal{B}$ to $F_\pi^2 \beta^r_{\pi\pi}$ respectively,
as defined in ref.~\cite{KMSF}.

In the above expressions, we have replaced the bare masses
by the physical masses in the (tadpole) logs and in the loop
functions $J^r$ and $M^r$.
One can check explicitly that there is no $\mu$ dependence 
in the above expression of the amplitude : 
for each polynomial in $s-2M_\pi^2,t-2M_\pi^2,u-2M_\pi^2$,
 the dependence of the LECs on the renormalisation scale $\mu$
cancels that of $J^r$ and $M^r$.

\subsection{$\pi K$ scattering amplitude}

We recall the expression obtained in ref.~\cite{BKM} for the $I=3/2$ amplitude:
\begin{eqnarray}
F_\pi^2 F_K^2 F^{3/2}_{\pi K} &=&
\frac{F_0^2}{6}\Bigg[
 2M_\pi^2+2M_K^2+\mo_\pi^2+\mo_K^2-3s\\
&& +\frac{\mu_\pi}{8}[66s-34M_\pi^2-54M_K^2-15\mo_\pi^2-21\mo_K^2] \nonumber\\
&& +\frac{\mu_K}{4}[30s-22M_\pi^2-18M_K^2-11\mo_\pi^2-9\mo_K^2] \nonumber\\
&& +\frac{\mu_\eta}{24}[54s-54M_\pi^2-18M_K^2-17\mo_\pi^2-11\mo_K^2] \nonumber\\
&& +8L_1^r (t-2M_\pi^2)(t-2M_K^2) + 4L_2^r
[(s-M_\pi^2-M_K^2)^2+(u-M_\pi^2-M_K^2)^2]\nonumber\\
&& +2L_3 [(u-M_\pi^2-M_K^2)^2+(t-2M_\pi^2)(t-2M_K^2)]\nonumber\\
&& +8L_4^r [\mo_\pi^2 (t-\frac{1}{2}s+\frac{1}{3}M_\pi^2-\frac{5}{3}M_K^2)
           +\mo_K^2 (t-s-\frac{4}{3}M_\pi^2+\frac{2}{3}M_K^2)]\nonumber\\
&& +\frac{4}{3} L_5^r [\mo_\pi^2 (2M_\pi^2-3s)+\mo_K^2 (2M_K^2-3s)]
   +\frac{8}{3} L_6^r [\mo_\pi^4 + 15 \mo_\pi^2 \mo_K^2 + 2 \mo_K^4]\nonumber\\
&& +\frac{8}{3} L_8^r [\mo_\pi^4 + 6 \mo_\pi^2 \mo_K^2 + \mo_K^4]\nonumber\\
&& +\frac{t}{2}(u-s) [M^r_{\pi\pi}(t)+\frac{1}{2}M^r_{KK}(t)]\nonumber\\
&& +\frac{3}{8}\{(s-t)[L_{\pi K}(u)-uM^r_{\pi K}(u)]+ (M_K^2-M_\pi^2)^2 M^r_{\pi K}(u)\}\nonumber\\
&& +\frac{3}{8}\{(s-t)[L_{K \eta}(u)-uM^r_{K \eta}(u)]+ (M_K^2-M_\pi^2)^2 M^r_{K \eta}(u)\}\nonumber\\
&& +\frac{1}{8} (M_K^2-M_\pi^2) K_{\pi K}(u) [5(u-M_\pi^2-M_K^2)+3\mo_\pi^2+3\mo_K^2]\nonumber\\
&& +\frac{1}{8} (M_K^2-M_\pi^2) K_{K \eta}(u) [3(u-M_\pi^2-M_K^2)+\mo_\pi^2+\mo_K^2]\nonumber\\
&& +\frac{1}{4} J^r_{\pi K}(s)(s-M_\pi^2-M_K^2)^2\nonumber\\
&& +\frac{1}{32} J^r_{\pi K}(u)[11(u-M_\pi^2-M_K^2)^2+10(u-M_\pi^2-M_K^2)(\mo_\pi^2+\mo_K^2)+3(\mo_\pi^2+\mo_K^2)^2]\nonumber\\
&& +\frac{1}{32} J^r_{K\eta}(u)[u-M_\pi^2-M_K^2+\frac{1}{3}(\mo_\pi^2+\mo_K^2)]^2\nonumber\\
&& +\frac{1}{8} J^r_{\pi\pi}(t)[4M_\pi^2-2t-3\mo_\pi^2][2M_K^2-t-2\mo_K^2]\nonumber\\
&& +\frac{3}{16} J^r_{KK}(t) [2M_\pi^2-t-2\mo_\pi^2][2M_K^2-t-2\mo_K^2]\nonumber\\
&& +\frac{1}{8} J^r_{\eta\eta}(t) \mo_\pi^2 [t-2M_K^2+\frac{10}{9}\mo_K^2]\nonumber
\end{eqnarray}
This expression is renormalisation-scale independent. In
both $\pi\pi$ and $\pi K$ scatterings, the one-loop expressions
obtained with the usual treatment of three-flavour
$\chi$PT~\cite{chpt3} are recovered if we
treat chiral series perturbatively
and neglect the (potentially large) difference between the truncated $O(p^2)$ 
expressions and the physical values of the pseudoscalar masses and 
decay constants.

\section{Computation of the amplitudes} \label{app:comp}

The amplitudes are smooth functions of the various experimental inputs. This
means in particular that there will be significant correlations among
the value of the same scattering amplitude at different points in the
Mandelstam plane. We compute these correlations according to the following procedure.
Let us call $a_k$ ($k=1\ldots n$) the parameters describing the variations of the
experimental inputs. To each of these parameters is 
attached an uncertainty ($\sigma_k$), and the correlations among them are 
encoded in a covariance matrix $D_{kl}$, or equivalently, a reduced 
covariance matrix $H_{kl}=D_{kl}/(\sigma_k \sigma_l)$.
We compute the mean value $m_i$ of the
observables $x_i$'s by setting all the parameters $a_k$ to their central value
$\bar{a}_k$: $m_i\equiv x_i(\bar{a}_k)$. Then, we 
vary the parameters one by one (the others being kept at their central
value) and compute each time:
\begin{equation}
  \Delta^k_i \equiv x_i\!\!\left(\bar{a}_k+\frac{\sigma_k}{\rho}\right) - m_i  = 
         \frac{\sigma_k}{\rho} \times \frac{\partial x_i}{\partial a_k} + \ldots
\end{equation}
where $\rho$ is a largish parameter (around 10), and the ellipsis denotes higher 
derivatives.
Once this is done for all the parameters, we compute the covariance matrix for the
observables:
\begin{equation} \label{gencorrmat}
  V_{ij} \equiv \rho^2 \sum_{kl}  \Delta^k_i \Delta^l_j H_{kl}
         = \sum_{kl} \frac{\partial x_i}{\partial a_k} \frac{\partial x_j}{\partial a_l} D_{kl} + \ldots
\end{equation}
The same procedure was followed in ref.~\cite{roypika}
to determine the correlation matrix between the two $\pi K$-scattering
lengths. 

For the $\pi\pi$ scattering amplitude, we obtain the following values and
errors for the amplitude at the limits of the subthreshold region $(s,t,u)$
\begin{eqnarray}
(2 M_\pi^2,M_\pi^2 , M_\pi^2 )  &\qquad & A=2.84  \pm 0.16 \\
(M_\pi^2/2,3/2 M_\pi^2,3/2 M_\pi^2) &\qquad & A=-1.03 \pm 0.12 \\
(M_\pi^2/2,3 M_\pi^2, M_\pi^2/2)    &\qquad & A=-1.08 \pm 0.11
\end{eqnarray}
For $\pi K$ scattering, we have in a similar way
\begin{eqnarray}
(M_K^2 ,2M_\pi^2,M_K^2)                     &\qquad & B=4.09 \pm 0.64 \qquad C=0 \\
(M_K^2,0,M_K^2+2M_\pi^2)                    &\qquad & B=2.96 \pm 0.60 \qquad C= -0.95  \pm 0.03 \\
(M_K^2+M_\pi^2,0,M_K^2+M_\pi^2)             &\qquad & B=2.61 \pm 0.60 \qquad C=0
\end{eqnarray}
The zeroes of $C$ are due to its antisymmetry under $s-u$ exchange. 
Uncertainties are correlated.

\section{Treatment of correlated data} \label{app:correl}

We expect strong correlations among the data points. This is reflected
by the fact that the matrix $C$ is nearly degenerate, and therefore cannot
be inverted easily. In order to treat this problem, one can diagonalize~\footnote{In practice, we use the singular value decomposition method described in ref.~\cite{NR},
which introduces two different rotation matrices on the left and on the right. This slight modification
does not alter the procedure outlined in this section.} the
matrix $C$:
\begin{equation}
C=UDU^T\,, \qquad D={\rm diag}(\lambda_1, \ldots, \lambda_n)\,, \qquad
UU^T=U^TU=1\,,
\end{equation}
which yields the corresponding likelihood:
\begin{eqnarray}
{\mathcal L}(\Delta V)
&=& \exp\left[-\frac{1}{2}\Delta V^T\ C^{-1}\ \Delta V\right]/\sqrt{\det (2\pi C)}\\
&=& \exp\left[-\frac{1}{2}\Delta V^T\ U D^{-1} U^T\ \Delta V\right]/\sqrt{\det (2\pi C)}\,.
\end{eqnarray}

Let us split the set of eigenvalues in two categories: large eigenvalues of
order 1, 
collected in the diagonal matrix $\tilde{D}$, and almost vanishing
eigenvalues, smaller than a cut-off and gathered in the diagonal matrix $D_0$:
\begin{equation}
D=\tilde{D}+D_0\,, \qquad \tilde{C}=U\tilde{D}U^T\,, \qquad C_0=UD_0U^T\,.
\end{equation}
 
The eigenvalues in $D_0$ are responsible for the near degeneracy of the
 matrix. In the corresponding directions, the exponential could be
 approximated with a Dirac distribution and would yield constrains on
 NNLO and higher-order remainders. Our approximation by a low-degree
 polynomial is expected to hold  at the level of a few percent: numerically,
 a perfect agreement between data and experiment occurs already if $\Delta
 V=O(1\%)$. Therefore, we cannot make much use of eigenvalues of the
 covariance matrix much smaller than $(1\%)^2=10^{-4}$. This leads us to
 limit the analysis to the subspace where $\tilde{D}$ is non-vanishing, and
 to define on this subspace $D^{-1}\equiv \tilde{D}^{-1}$ (see Ch.~2.6 in
 ref.~\cite{NR} for a more detailed discussion on the relationships between 
 singular value decomposition and matrix inversion).
 We chose to set the limit between small and large eigenvalues of order
 $10^{-8}$ (with only a very mild dependence of our results on the exact value
 of the cutoff).

\section{Impact of combining $\pi\pi$ and $\pi K$ data for 
marginal CL profiles} \label{app:comb}

As shown in sec.~\ref{sec:rescls}, the analysis of $\pi\pi$ data in our
framework puts a lower bound on $r$ ($r\geq 12$ at 68~\% CL), and an upper
bound on $X(3)$ and $Z(3)$ (below 0.85 and 0.95 respectively). 
$\pi K$ scattering does seem to bring only a lower bound on $X(3)$. The
combination of these two pieces of information proves much more powerful and
allows one to extract CL intervals on $r$, $X(3)$, $Y(3)$ and $Z(3)$.
Indeed, we recall that our statistical method, inspired by the Rfit
approach~\cite{rfit}, consists in the following steps : determine the absolute minimum of
$\chi^2$ first, then fix a particular theoretical parameter $a$ (among $r$, $X(3)$, $Y(3)$, $Z(3)$)
and compute the corresponding relative 
minimum of $\chi^2$, finally extract a CL (actually a $P$-value~\cite{statlect} 
from the difference 
between the two values of $\chi^2$. This amounts to computing 
\begin{equation}
{\rm CL}[a|{\rm data}]={\rm Max}_\mu {\rm CL}[a;\mu|{\rm data}]
\end{equation}
where $\mu$ collects all the remaining theoretical parameters 
(including $L_i$ and NNLO remainders). Therefore, the CL profiles obtained in
our approach correspond to upper bounds on the CLs. In particular, it is enough that
one set of theoretical parameters $\mu$ yields the same value of $\chi^2$ as
the absolute minimum to get ${\rm CL}(a)=1$.

We have many theoretical parameters for the description of the scattering
amplitudes : in addition to $r$, $X(3)$ and $Z(3)$, we have the $O(p^4)$ LECs
$L_{1,2,3}$ and 
(direct and indirect) NNLO remainders. Therefore, it is not particularly
surprising that either $\pi\pi$ or $\pi K$ scattering alone is not enough to
simultaneously put constrains on all these parameters : many equivalent
situations (with identical $\chi^2$ and thus ${\rm CL}$) 
can be obtained with different sets
of theoretical parameters. This degeneracy, in particular for the minimum
$\chi^2$, comes from the possibility 
of compensating a variation in $a$ by a modification 
of the remaining theoretical parameters $\mu$ within the
allowed ranges, which tends to yield  
flat CL profiles when we consider only one amplitude.

This underdetermination of the theoretical parameters -- and the resulting
degeneracy in CL values -- is lifted
once several sets of different sources are considered. In the
present case, $\pi\pi$ data puts constraints on the quark mass ratio $r$
and on some of the $O(p^4)$ derivative couplings. Because of these
constraints, the regions of theoretical parameters with identical CLs are
reduced, and thus the CL curves associated with $\pi K$ scattering exhibit
more distinctive features.

\begin{figure}
\begin{center}

\includegraphics[width=11cm]{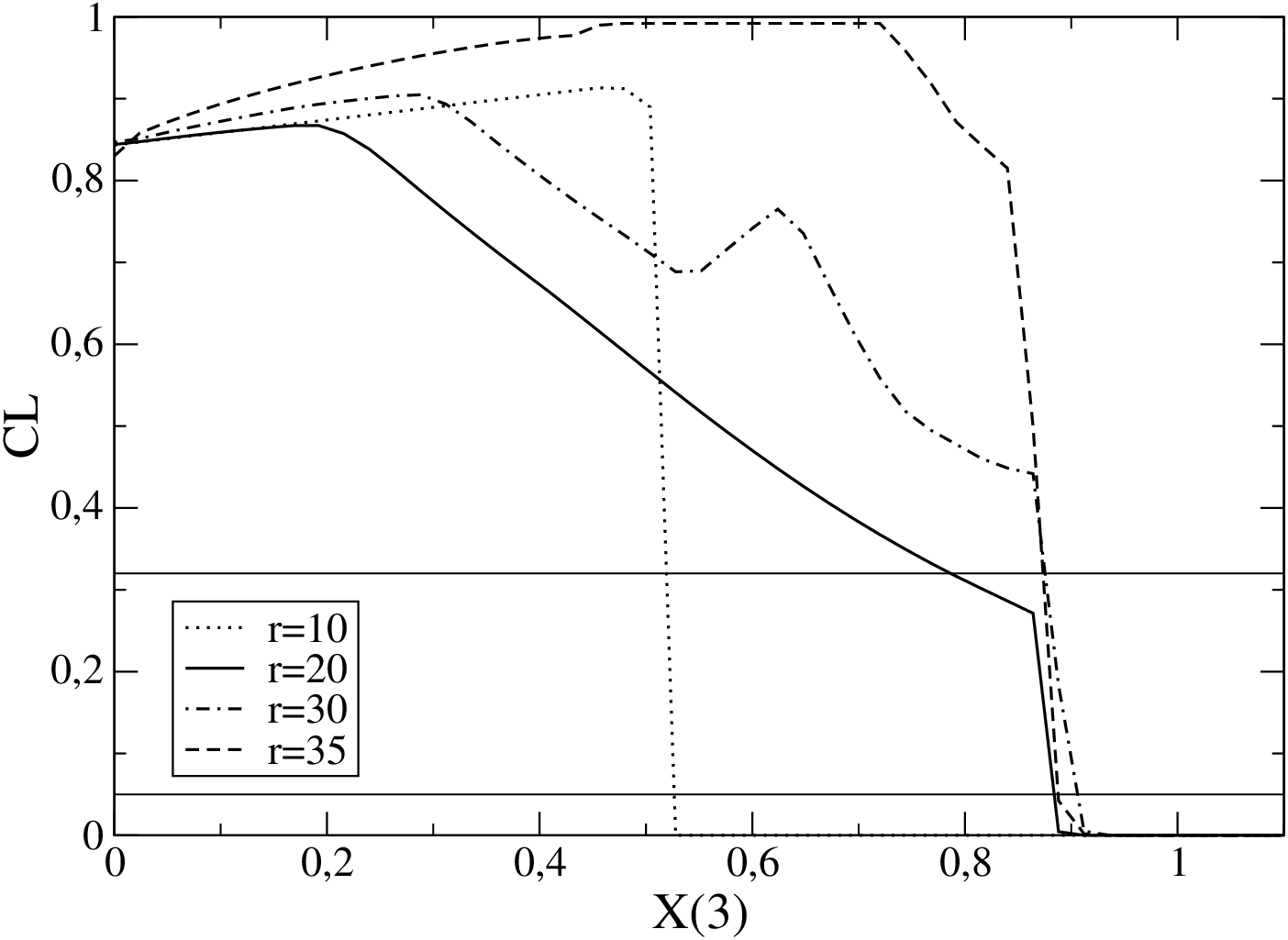}

\vspace{1.3cm}

\includegraphics[width=11cm]{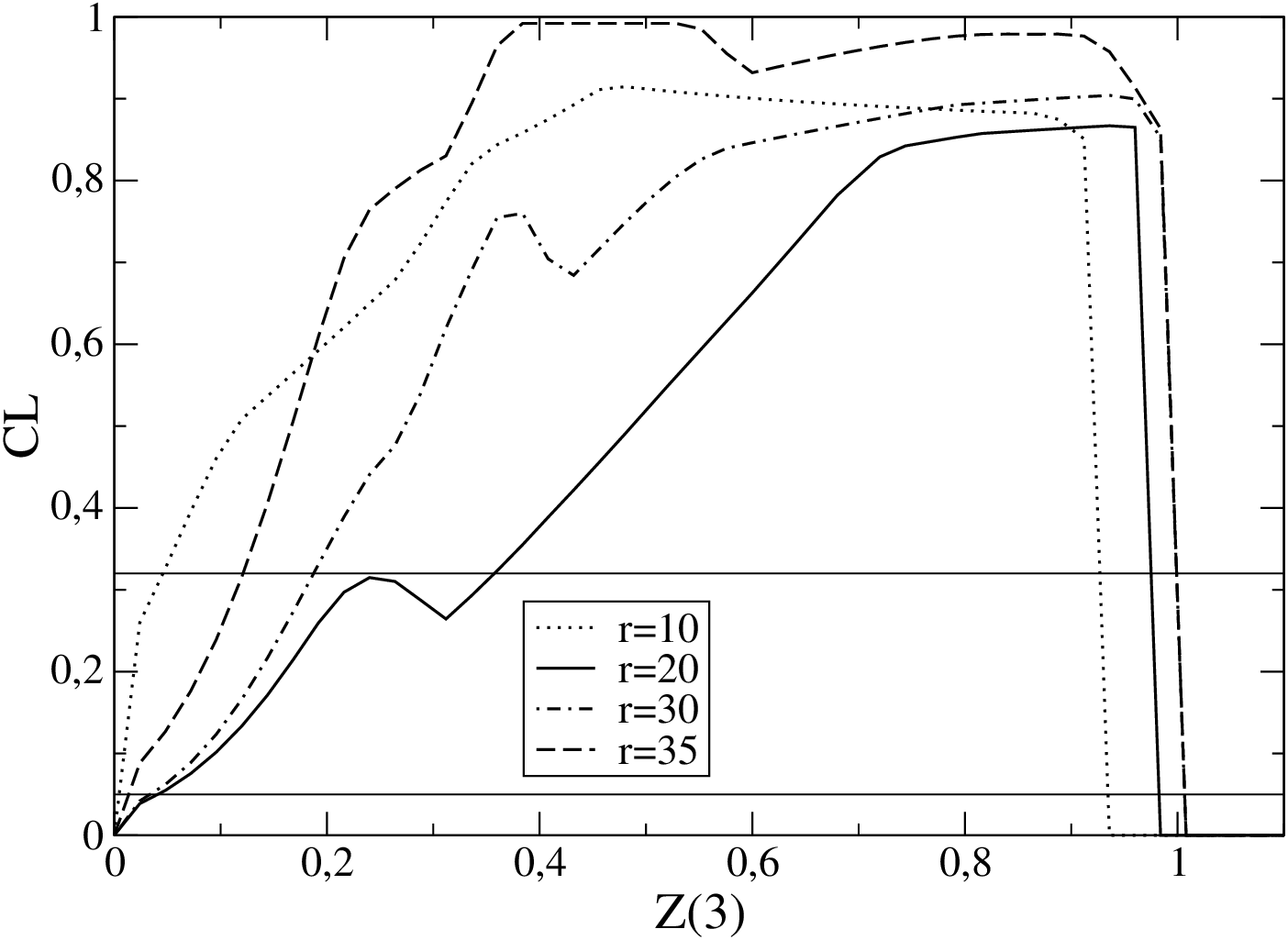}

\caption{CL profiles for $X(3)=2m\Sigma(3)/(F_\pi^2 M_\pi^2$ (top) and
  $Z(3)=F_0^2/F_\pi^2$ (bottom), obtained from the combination of
  the experimental information on $\pi\pi$ and $\pi K$ scatterings, with
  $L_1,L_2,L_3$ set to the values minimizing the complete $\chi^2$ (see text). The three curves
  corresponding to four different (imposed) values of $r=10,20,30,35$. The two
  horizontal lines indicate the confidence intervals at 68 and 95\% CL.}

\label{fig:rlfix}
\end{center}
\end{figure}

As an illustration of this phenomenon, we compute the CL for $\pi K$ data
alone with $L_1,L_2,L_3$ and $r$ fixed to specific values, 
in order to mimic the interplay between 
$\pi\pi$ and $\pi K$ data in CL curves. Let us set $L_1,L_2,L_3$ 
to the values corresponding to the absolute minimum of the $\chi^2$
when both $\pi\pi$ and $\pi K$ data are considered
\begin{equation}
L_1^r(M_\rho)=-0.31 \cdot 10^{-3}\,, \qquad
L_2^r(M_\rho)= 2.12 \cdot 10^{-3}\,, \qquad
L_3^r(M_\rho)=-0.64\cdot 10^{-3}\,,
\end{equation}
and let us set $r$ to four different values $r=10,20,30,35$. With these
theoretical parameters fixed, we determine ${\rm CL}[X(3);r,L_{1,2,3}|\pi K]$ 
and ${\rm CL}[Z(3);r,L_{1,2,3}|\pi K]$
profiles, which are drawn in fig.~\ref{fig:rlfix}. 
We can see that the CL profiles of $X(3)$ and $Z(3)$
for $r=20$ are very similar to the solid lines shown in figs.~\ref{fig:rx}
and \ref{fig:yz}, corresponding to $CL[X(3)|\pi\pi,\pi K]$
and $CL[Z(3)|\pi\pi,\pi K]$. On the other hand, the curves 
for $r=10,30,35$ are somewhat broader and flatter. 

The CL curves obtained for $\pi K$ 
scattering in figs.~\ref{fig:rx}
and \ref{fig:yz} are ${\rm CL}[X(3)|\pi K]$ and  
${\rm CL}[Z(3)|\pi K]$, which 
correspond to the envelope of all the CL profiles of the form
${\rm CL}[X(3);r,L_{1,2,3}|\pi K]$ and ${\rm CL}[Z(3);r,L_{1,2,3}|\pi K]$
when varying $r$ (and  $L_{1,2,3}$). This superimposition, dominated by
values of $r$ around 35, eventually yields the flat profiles in $X(3)$
and $Z(3)$ observed in figs.~\ref{fig:rx}
and \ref{fig:yz}. On the other hand, when we combine $\pi\pi$ and $\pi K$
scatterings, a value of $r$ around 20 is preferred by $\pi\pi$ data (together
with some ranges for $L_{1,2,3}$). Therefore,
the contribution from $\pi K$ scattering to ${\rm CL}[X(3)|\pi\pi,\pi K]$
and ${\rm CL}[Z(3)|\pi\pi,\pi K]$ is close to the CL curve obtained 
for ${\rm CL}[X(3);r=20,L_{1,2,3}|\pi K]$ and 
${\rm CL}[Z(3);r=20,L_{1,2,3}|\pi K]$, i.e. $r=20$ in fig.~\ref{fig:rlfix}.
This phenomenon explains how the information on theoretical parameters
can be hidden in $\pi K$ scattering, but is unveiled 
once combined with $\pi\pi$ scattering, lifting the degeneracy in the CLs
and leading to a sharper  determination of $X(3)$ and $Z(3)$.

\begin{thebibliography}{99}

\bibitem{param}
S.~Descotes-Genon, L.~Girlanda and J.~Stern,
JHEP {\bf 0001}, 041 (2000)
[\arxiv{hep-ph/9910537}].

\bibitem{E865}
S.~Pislak {\it et al.}  [BNL-E865 Collaboration],
Phys. Rev. Lett. 87 (2001) 221801
[\arxiv{hep-ex/0106071}];
Phys. Rev. D 67 (2003) 072004
[\arxiv{hep-ex/0301040}].

\bibitem{ACGL}
B.~Ananthanarayan, G.~Colangelo, J.~Gasser and H.~Leutwyler,
Phys.\ Rept.\  {\bf 353} (2001) 207
[\arxiv{hep-ph/0005297}].

\bibitem{pipi}
S.~Descotes-Genon, N.~H.~Fuchs, L.~Girlanda and J.~Stern,
Eur.\ Phys.\ J.\ C {\bf 24}, 469 (2002)
[\arxiv{hep-ph/0112088}].

\bibitem{CGL}
G.~Colangelo, J.~Gasser and H.~Leutwyler,
Phys.\ Rev.\ Lett.\  {\bf 86} (2001) 5008
[\arxiv{hep-ph/0103063}].

\bibitem{chpt2}
J.~Gasser and H.~Leutwyler,
Annals Phys.\  {\bf 158}, 142 (1984).

\bibitem{NA48}
L.~Masetti [NA48 collaboration],
Talk at the 33rd International Conference on High Energy Physics (ICHEP 06),
  \arxiv{hep-ex/0610071}.


\bibitem{lattice}
  L.~Del Debbio, L.~Giusti, M.~Luscher, R.~Petronzio and N.~Tantalo,
  \arxiv{hep-lat/0610059},
  \arxiv{hep-lat/0701009}.

D.~Becirevic, Ph.~Boucaud, V.~Lubicz, G.~Martinelli, F.~Mescia, S.~Simula and C.~Tarantino,
  Phys.\ Rev.\  D {\bf 74}, 034501 (2006)
  [\arxiv{hep-lat/0605006}].

Ph.~Boucaud {\it et al.}  [ETM Collaboration],
  \arxiv{hep-lat/0701012}.

\bibitem{chpt3}
J.~Gasser and H.~Leutwyler,
Nucl.\ Phys.\ B {\bf 250}, 465 (1985).

\bibitem{roypika}
P.~B\"uttiker, S.~Descotes-Genon and B.~Moussallam,
Eur.\ Phys.\ J.\ C {\bf 33}, 409 (2004)
[\arxiv{hep-ph/0310283}].

\bibitem{ordfluc}
S.~Descotes-Genon, L.~Girlanda and J.~Stern,
Eur.\ Phys.\ J.\ C {\bf 27}, 115 (2003)
[\arxiv{hep-ph/0207337}].

\bibitem{resum}
S.~Descotes-Genon, N.~H.~Fuchs, L.~Girlanda and J.~Stern,
Eur.\ Phys.\ J.\ C {\bf 34}, 201 (2004)
[\arxiv{hep-ph/0311120}].


\bibitem{rfit}
 A.~Hocker, H.~Lacker, S.~Laplace and F.~Le Diberder,
  Eur.\ Phys.\ J.\ C {\bf 21} (2001) 225
  [\arxiv{hep-ph/0104062}].

\bibitem{statlect}
W.J. Metzger, \emph{Statistical Methods in Data Analysis},\\
\arxiv{http://www.hef.kun.nl/~wes/stat\_course/statist.pdf}

F. James, \emph{Statistical Methods in experimental physics},
World Scientific 2006.

\bibitem{gchpt}
N.~H.~Fuchs, H.~Sazdjian and J.~Stern,
  Phys.\ Lett.\  B {\bf 238} (1990) 380,
  Phys.\ Lett.\  B {\bf 269} (1991) 183,
  Phys.\ Rev.\  D {\bf 47} (1993) 3814
  [\arxiv{hep-ph/9301244}].

\bibitem{KMSF}
M.~Knecht, B.~Moussallam, J.~Stern and N.~H.~Fuchs,
  Nucl.\ Phys.\ B {\bf 457}, 513 (1995)
  [\arxiv{hep-ph/9507319}].

\bibitem{dispff}
B.~Moussallam,
Eur.\ Phys.\ J.\ C {\bf 14}, 111 (2000)
[\arxiv{hep-ph/9909292}] ;
JHEP {\bf 0008}, 005 (2000)
[\arxiv{hep-ph/0005245}].

\bibitem{dispzr}
S.~Descotes-Genon,
JHEP {\bf 0103}, 002 (2001)
[\arxiv{hep-ph/0012221}].

\bibitem{twoloopscal}
J.~Bijnens and P.~Dhonte,
JHEP {\bf 0310}, 061 (2003)
[\arxiv{hep-ph/0307044}].

\bibitem{twolooppipi}
J.~Bijnens, P.~Dhonte and P.~Talavera,
JHEP {\bf 0401}, 050 (2004)
[\arxiv{hep-ph/0401039}].

\bibitem{twolooppika}
J.~Bijnens, P.~Dhonte and P.~Talavera,
JHEP {\bf 0405}, 036 (2004)
[\arxiv{hep-ph/0404150}].

\bibitem{twolooprev}
J.~Bijnens,
Prog.\ Part.\ Nucl.\ Phys.\  {\bf 58} (2007) 521
  [\arxiv{hep-ph/0604043}].

\bibitem{twoloopmass}

  G.~Amoros, J.~Bijnens and P.~Talavera,
  Nucl.\ Phys.\  B {\bf 568} (2000) 319
  [\arxiv{hep-ph/9907264}];
  Nucl.\ Phys.\  B {\bf 602} (2001) 87
  [\arxiv{hep-ph/0101127}].

\bibitem{twoloopkl3}
  J.~Bijnens and P.~Talavera,
  Nucl.\ Phys.\  B {\bf 669} (2003) 341
  [\arxiv{hep-ph/0303103}].

\bibitem{zr}
S.~Descotes-Genon and J.~Stern,
Phys.\ Lett.\ B {\bf 488}, 274 (2000)
[\arxiv{hep-ph/0007082}].

\bibitem{BKM}
  V.~Bernard, N.~Kaiser and U.~G.~Meissner,
  Phys.\ Rev.\ D {\bf 43} (1991) 2757.

\bibitem{lecsize}
 A.~Manohar and H.~Georgi,
  Nucl.\ Phys.\  B {\bf 234} (1984) 189.

H.~Georgi,
  Phys.\ Lett.\  B {\bf 298} (1993) 187
  [\arxiv{hep-ph/9207278}].

\bibitem{ckmfitter}
J.~Charles {\it et al.}  [CKMfitter Group],
  Eur.\ Phys.\ J.\ C {\bf 41} (2005) 1
  [\arxiv{hep-ph/0406184}].


\bibitem{nfdep}
J.~Stern,
  \arxiv{hep-ph/9801282}.

S.~Descotes-Genon and J.~Stern,
  Phys.\ Rev.\  D {\bf 62}, 054011 (2000)
  [\arxiv{hep-ph/9912234}].

\bibitem{naturalp6}
K.~Kampf and B.~Moussallam,
  Eur.\ Phys.\ J.\  C {\bf 47} (2006) 723
  [\arxiv{hep-ph/0604125}].

\bibitem{satur}
 G.~Ecker, J.~Gasser, A.~Pich and E.~de Rafael,
  Nucl.\ Phys.\  B {\bf 321} (1989) 311.

\bibitem{amoros}
G.~Amoros, J.~Bijnens and P.~Talavera,
Phys.\ Lett.\ B {\bf 480} (2000) 71
[\arxiv{hep-ph/9912398}],
Nucl.\ Phys.\ B {\bf 585} (2000) 293
[Erratum-ibid.\ B {\bf 598} (2001) 665]
[\arxiv{hep-ph/0003258}].


\bibitem{MILC}
  C.~Bernard {\it et al.}  [MILC Collaboration],
  \arxiv{hep-lat/0609053}.

\bibitem{staggered}
M.~Creutz,
  \arxiv{hep-lat/0603020},
  \arxiv{hep-lat/0701018}.

C.~Bernard, M.~Golterman, Y.~Shamir and S.~R.~Sharpe,
  \arxiv{hep-lat/0603027}.

S.~Durr,
  PoS {\bf LAT2005} (2006) 021
  [\arxiv{hep-lat/0509026}].

S.~R.~Sharpe,
  PoS {\bf LAT2006} (2006) 022
  [\arxiv{hep-lat/0610094}].

\bibitem{SChPT}
  S.~R.~Sharpe and R.~S.~Van de Water,
  Phys.\ Rev.\  D {\bf 71} (2005) 114505
  [\arxiv{hep-lat/0409018}].

\bibitem{mixed}
 J.~W.~Chen, D.~O'Connell, R.~S.~Van de Water and A.~Walker-Loud,
  Phys.\ Rev.\  D {\bf 73} (2006) 074510
  [\arxiv{hep-lat/0510024}].

D.~O'Connell,
  \arxiv{hep-lat/0609046}.

 J.~W.~Chen, D.~O'Connell and A.~Walker-Loud,
  Phys.\ Rev.\  D {\bf 75} (2007) 054501
  [\arxiv{hep-lat/0611003}].

\bibitem{lattest}
  S.~Descotes-Genon,
  Eur.\ Phys.\ J.\  C {\bf 40} (2005) 81
  [\arxiv{hep-ph/0410233}].

\bibitem{expercusp}
J.~R.~Batley {\it et al.}  [NA48/2 Collaboration],
  Phys.\ Lett.\  B {\bf 633} (2006) 173
  [\arxiv{hep-ex/0511056}].

\bibitem{cusp}
N.~Cabibbo,
  Phys.\ Rev.\ Lett.\  {\bf 93} (2004) 121801
  [\arxiv{hep-ph/0405001}].

N.~Cabibbo and G.~Isidori,
  JHEP {\bf 0503} (2005) 021
  [\arxiv{hep-ph/0502130}].
  
G.~Colangelo, J.~Gasser, B.~Kubis and A.~Rusetsky,
  Phys.\ Lett.\  B {\bf 638} (2006) 187
  [\arxiv{hep-ph/0604084}].


E.~Gamiz, J.~Prades and I.~Scimemi,
  Eur.\ Phys.\ J.\  C {\bf 50} (2007) 405
  [\arxiv{hep-ph/0602023}].

 
\bibitem{experdl4}
J.~M.~Link {\it et al.}  [FOCUS Collaboration],
  Phys.\ Lett.\  B {\bf 607} (2005) 67
  [\arxiv{hep-ex/0410067}].

  M.~R.~Shepherd {\it et al.}  [CLEO Collaboration],
  Phys.\ Rev.\  D {\bf 74}, 052001 (2006)
  [\arxiv{hep-ex/0606010}].

\bibitem{dl4}
C.~L.~Y.~Lee, M.~Lu and M.~B.~Wise,
  Phys.\ Rev.\  D {\bf 46}, 5040 (1992).

B.~Ananthanarayan and K.~Shivaraj,
  Phys.\ Lett.\  B {\bf 628} (2005) 223
  [\arxiv{hep-ph/0508116}].

\bibitem{expertau}
R.~Barate {\it et al.}  [ALEPH Collaboration],
  Eur.\ Phys.\ J.\  C {\bf 11} (1999) 599
  [\arxiv{hep-ex/9903015}].

G.~Abbiendi {\it et al.}  [OPAL Collaboration],
  Eur.\ Phys.\ J.\  C {\bf 35} (2004) 437
  [\arxiv{hep-ex/0406007}].

\bibitem{tau}
  M.~Jamin, A.~Pich and J.~Portoles,
  Phys.\ Lett.\  B {\bf 640}, 176 (2006)
  [\arxiv{hep-ph/0605096}].

  M.~Jamin, J.~A.~Oller and A.~Pich,
  Phys.\ Rev.\  D {\bf 74}, 074009 (2006)
  [\arxiv{hep-ph/0605095}].

\bibitem{NR}
W.H. Press, S.A. Teukolsky, W.T. Vetterling, B.P. Flannery,
\emph{Numerical recipes - The art of scientific computing}, 
Cambridge University Press.



\end{thebibliography}
\end{document}